\documentclass{article}\usepackage{amssymb}
\usepackage{graphicx} 
\usepackage{hyperref}
\usepackage{caption}
\usepackage{subcaption}
\usepackage{geometry}
\usepackage{authblk}
\usepackage{amsmath}
\usepackage{multirow}
\usepackage{xcolor}
\usepackage{booktabs}
\usepackage{siunitx}

\usepackage{listings}
\definecolor{codegreen}{rgb}{0,0.6,0}
\definecolor{codegray}{rgb}{0.5,0.5,0.5}
\definecolor{codepurple}{rgb}{0.58,0,0.82}
\definecolor{backcolour}{rgb}{0.95,0.95,0.92}

\usepackage{mathtools}
\DeclarePairedDelimiter\bra{\langle}{\rvert}
\DeclarePairedDelimiter\ket{\lvert}{\rangle}
\DeclarePairedDelimiterX\braket[2]{\langle}{\rangle}{#1\,\delimsize\vert\,\mathopen{}#2}

\title{Improving thermal state preparation of Sachdev-Ye-Kitaev model with reinforcement learning {on quantum hardware}}
\author{Akash Kundu\footnote{Corresponding author, \href{mailto:akash.kundu@helsinki.fi}{akash.kundu@helsinki.fi}}}
\affil{QTF Centre of Excellence, Department of Physics, University of Helsinki, Finland}
\date{}

\begin{document}

\maketitle

\begin{abstract}
    The Sachdev-Ye-Kitaev (SYK) model, known for its strong quantum correlations and chaotic behavior, serves as a key platform for quantum gravity studies. However, variationally preparing thermal states on near-term quantum processors for large systems ($N>12$, where $N$ is the number of Majorana fermions) presents a significant challenge due to the rapid growth in the complexity of parameterized quantum circuits. This paper addresses this challenge by integrating reinforcement learning (RL) with convolutional neural networks, employing an iterative approach to optimize the quantum circuit and its parameters. The refinement process is guided by a composite reward signal derived from entropy and the expectation values of the SYK Hamiltonian. This approach reduces the number of CNOT gates by two orders of magnitude for systems $N\geq12$ compared to traditional methods like first-order Trotterization. We demonstrate the effectiveness of the RL framework in both noiseless and noisy quantum hardware environments, maintaining high accuracy in thermal state preparation. This work advances a scalable, RL-based framework with applications for quantum gravity studies and out-of-time-ordered thermal correlators computation in quantum many-body systems on near-term quantum hardware. 
    {The code is available at~\href{https://github.com/Aqasch/solving_SYK_model_with_RL}{\textcolor{blue}{solving\_SYK\_with\_RL}} repository}.
\end{abstract}

\section{Introduction}

The holographic duality~\cite{witten1998anti} establishes a profound connection between a specific class of quantum field theories in $d$ and quantum gravity in $d+1$ dimensions. This strong/weak duality allows the exploration of strongly coupled field theories via classical supergravity and vice versa. Despite its power, there are no known cases where both sides of the duality can be simultaneously and analytically solved. The Sachdev-Ye-Kitaev (SYK) model, a simplified variant of the Sachdev-Ye model~\cite{sachdev1993gapless}, was introduced in a series of seminal talks by Kitaev~\cite{kitaev2015simple,kitaev2015simplep2}. This model exhibits holographic properties and is particularly amenable to study in the strong-coupling limit.

The SYK model describes a system of $N$ Majorana fermions in a quantum mechanical setting (i.e., in $0+1$ dimensions), where the interaction involves random couplings among $q$ distinct fermions. These random couplings, denoted by $J_{i_1 i_2 \ldots i_q}$, are independent Gaussian variables with zero mean and variance given by
\[
\langle J_{i_1 i_2 \ldots i_q}^2 \rangle = \frac{J^2 (q-1)!}{N^{q-1}}.
\]
In the regime where $N$ is very large and the temperature is low ($N \gg \beta J \gg 1$, with $\beta$ the inverse temperature and $J$ the characteristic interaction strength), the model displays an approximate conformal symmetry and is related to near-extremal black holes with an nAdS$_2$ (near-AdS$_2$) geometry. Meanwhile, the SYK model saturates the chaos bound~\cite{maldacena2016bound}, a key feature of holographic systems. Moreover, SYK model is a prominent example of a fast scrambler~\cite{kim2021dirac}, known for its logarithmic growth of complexity and its saturation of the bounds proposed by the Brown-Susskind conjecture regarding quantum information scrambling in black holes~\cite{brown2017quantum}. Its computational tractability in $0+1$ dimensions has enabled extensive numerical studies, with systems comprising up to 60 Majorana fermions analyzed~\cite{gur2018does, kobrin2021many}. Over the past decade, the SYK model has been extensively studied in both condensed matter and high-energy physics~\cite{maldacena2016remarks, polchinski2016spectrum, gross2017generalization}. For a comprehensive review, see refs.~\cite{chowdhury2022sachdevreview, sarosi2017adsreview, rosenhaus2019introductionreview}.

Owing to its pivotal role in holographic duality, rigorous analysis of both vacuum and finite-temperature states in the SYK model becomes essential. The ground state displays extensive entanglement entropy scaling linearly with system size, with coefficients explicitly computed in Ref.~\cite{huang2019eigenstate} – a property making it classically intractable to simulate \cite{passetti2023can}. A hallmark characteristic emerges in thermal correlation functions: out-of-time-order correlators (OTOCs) exhibit maximal quantum Lyapunov exponents $\lambda_L = 2\pi/\beta$ at low temperatures ($\beta J \to \infty$) in the large-$N$ limit \cite{maldacena2016bound}. This chaotic behavior necessitates precise preparation of thermal states, a key objective of this work. While prior studies have probed low-energy physics through classical simulation, we propose leveraging variational quantum algorithms \cite{cerezo2021variational, mcclean2016theory} to overcome these limitations and enable quantum simulations of nonperturbative dynamics.
\begin{figure}[h!]
    \centering
\includegraphics[width=\linewidth]{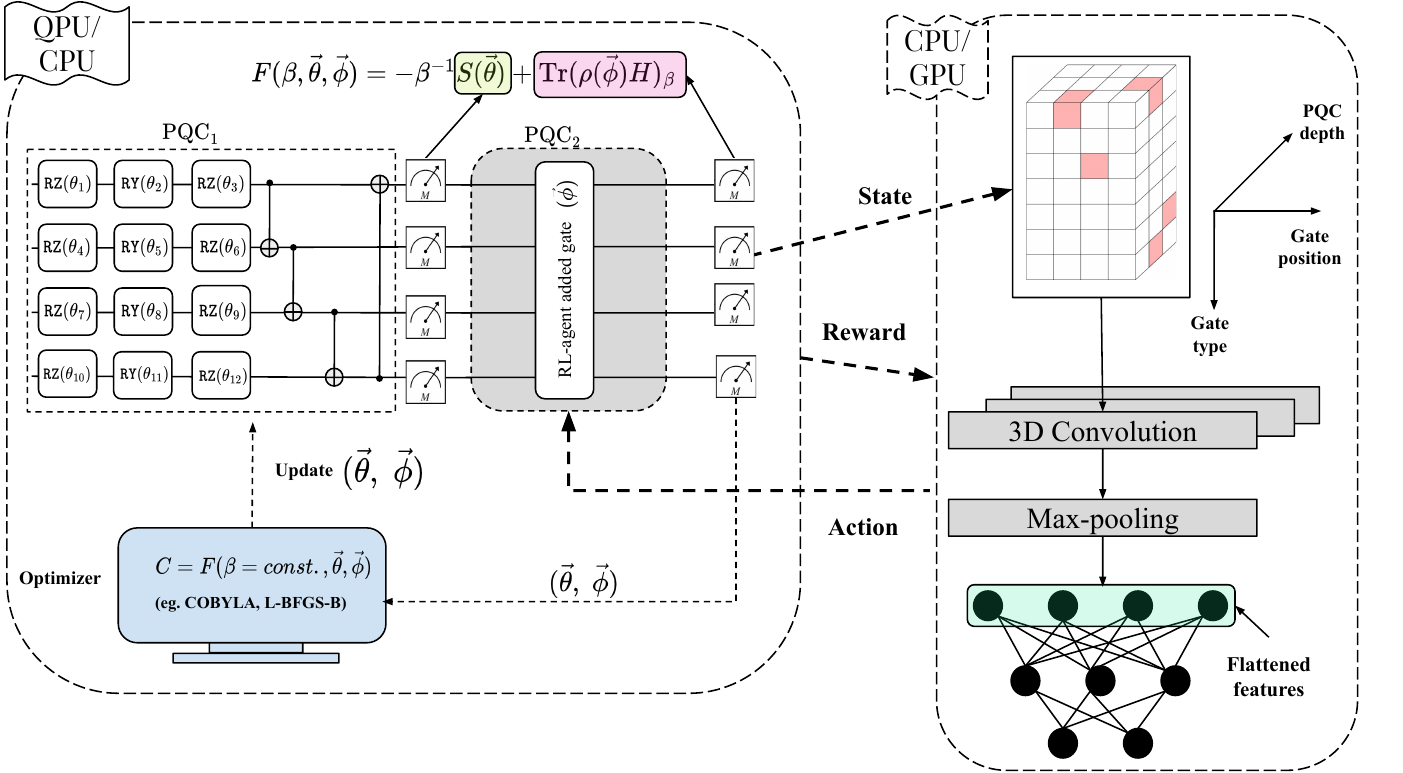}
    \caption{\small\textbf{Illustration of the main algorithm, where the environment, i.e., the PQC$_2$ in the variational framework encoded in a 3D tensor, interacts with an agent that is constructed using a 3D convolutional neural network under a reinforcement learning (RL) framework}. A classical optimizer optimizes the variational circuits and provides a reward signal to guide the agent in selecting subsequent actions.
    Upon receiving a reward and the current state $s$ of the environment, the RL-agent chooses an action $a$ in $s$ in accordance with a $\epsilon$-greedy policy. Using this framework, in Fig.~\ref{fig:cnot_resources} we show that the number of CNOT gates is reduced by $10^2$-fold beyond $N=14$ Majorana fermions in PQC$_2$ compared to a first-order Trotter approximation.}
    \label{fig:main_algorithm}
\end{figure}

{While condensed matter systems, such as topological interfaces~\cite{pikulin2017black}-naturally realize large-$N$ SYK-like physics, and classical approaches like exact diagonalization can currently handle up $N=34$ to 0 + 1-dimensional models and have
been studied up to $N=60$ Majorana fermions~\cite{kobrin2021many}, we see quantum simulation as a promising alternative route. Instead of trying to outdo classical computation, our focus is on harnessing the unique strengths of quantum hardware, like precise control over parameters and direct access to non-equilibrium dynamics~\cite{cao2020towards, luo2019quantum}. By developing efficient state preparation techniques tailored for near-term devices, we aim to explore regimes that are out of reach for ED, such as open-system dynamics or much larger $N$. In this way, our approach complements classical methods and sets the stage for tackling high-$N$ problems that could reveal new physics as quantum technology continues to advance.}

Recent efforts have applied various quantum simulation techniques to the SYK model, including digital simulation~\cite{garcia2017digital}, teleportation protocols~\cite{lykken2024long}, bosonic correlation computation~\cite{luo2019quantum}, and the preparation of the ground state of the coupled SYK model~\cite{su2020variational, kim2021entanglement}. Notably, ref.~\cite{araz2024thermal} employed a variational quantum algorithm to prepare thermal states of dense and sparse SYK models for $6 \leq N \leq 12$ Majorana fermions across a range of temperatures. 
Extending the framework developed in Ref.~\cite{asaduzzaman2023model}, this work investigates quantum hardware implementations for probing real-time evolution in the SYK model.
However, this approach faces significant challenges, including the rapid growth in the number of parameters and controlled-NOT (CNOT) gates with increasing $N$. These limitations restricted simulations to $N \leq 12$, as larger systems incurred prohibitive computational overhead.
\begin{figure}[h!]
    \centering
    \includegraphics[width=\linewidth]{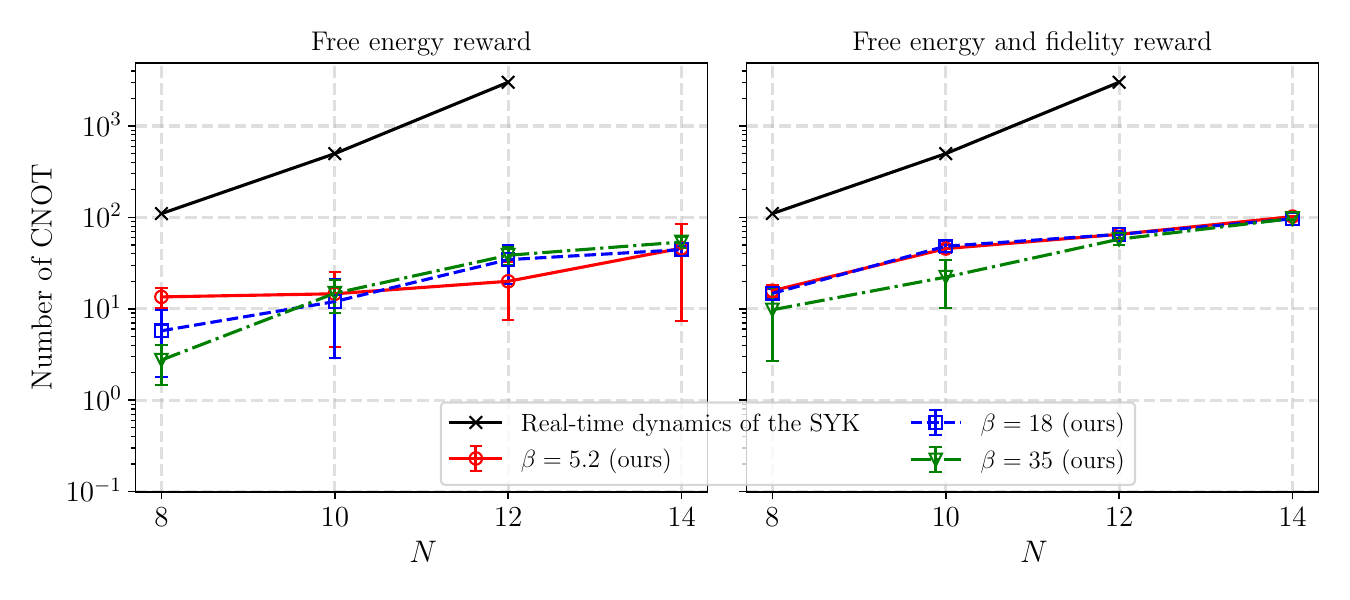}
    \caption{\small\textbf{The CNOT gates in the PQC$_2$ are $10^2$-fold improved beyond $N=12$ if the thermal state is prepared in the reinforcement learning framework compared to the real-time dynamics obtained by a first-order Trotterization~\cite{asaduzzaman2023model}}. The resource count is obtained in a noiseless scenario where both the environment and the agent are noiseless. Using the reward signal presented in Eq.~\ref{case:error_rwd} (where the signal constitutes the error in free energy, hence ``Free energy reward") and in Eq.~\ref{case:error_and_fidelity_rwd} (where the signal constitutes the error in free energy and the thermal state fidelity, hence the ``Free energy and fidelity reward") and for different temperatures ($1/\beta$). The ``\textit{ours}" denotes the variational thermal state preparation under the reinforcement learning framework in Fig.~\ref{fig:main_algorithm}. {In Appendix~\ref{appendix:scaling_of_snot_analysis} we further analyze the scaling of CNOT with the number of qubits and show that for our method the CNOT scales polynomially.}}
    \label{fig:cnot_resources}
\end{figure}

In this work, we address these challenges by integrating reinforcement learning (RL) with convolutional neural networks (CNNs)~\cite{goodfellow2016deep} into the variational quantum algorithm (VQA) framework. RL has recently emerged as a powerful tool for addressing quantum computing problems on near-term quantum devices~\cite{patel2024curriculum,kundu2024enhancing,patel2024reinforcement,ostaszewski2021reinforcement,kuo2021quantum,moro2021quantum,erdman2024artificially}. In this approach, quantum circuits are modelled as sequences of actions governed by a trainable policy. The VQA cost function, independently optimized using a classical optimizer, provides periodic feedback that contributes to the final reward signal. This reward signal updates the policy, maximizing expected returns and guiding the selection of optimal actions for subsequent steps. For a brief review of RL for quantum circuit optimization check ref.~\cite{kundu2024reinforcementthesis}.

State-of-the-art RL-based approaches typically update the policy using the reward signal only after the VQA optimization is complete. In contrast, our algorithm, as illustrated in Fig.~\ref{fig:main_algorithm}, employs a composite reward function updated iteratively. This reward function combines two key components: the entropy measured from the output of the first parameterized quantum circuit (PQC$_1$)\footnote{PQC$_1$ consists of a layer of \texttt{RZ}-\texttt{RY}-\texttt{RZ} gates on all qubits and all-to-all connected CNOT gates.}, and the expectation value of the SYK Hamiltonian computed after applying the second parameterized quantum circuit (PQC$_2$). Moreover, instead of frequently used feedforward neural networks (FNN)~\cite{sazli2006brief,svozil1997introduction} we utilize a 3D-CNN to enhance the trainability of the RL framework.

In essence, we train a CNN within a reinforcement learning framework using an $\epsilon$-greedy policy guided by this composite reward signal in two different scenarios: (1) environment: noiseless, agent: noiseless up to $N=14$ Majorana fermions; and (2) environment: noisy with constrained qubit connectivity, agent: noiseless for $N=8$ Majorana fermions. The RL framework enables the identification of optimal parameterized quantum circuits for preparing thermal states across various temperatures, significantly reducing computational overhead. To demonstrate the effectiveness of the framework in Fig.~\ref{fig:cnot_resources} we summarize one of the main outcomes of the paper in scenario (1), which says that in a noiseless scenario, \textit{compared to parameterized circuits obtained using real-time dynamics, i.e., the first-order Trotterization of the SYK Hamiltonian, our method (depicted in Fig.~\ref{fig:main_algorithm}) reduces the number of CNOT gates at least by a factor of $10^2$ when $N\geq12$} while maintaining high accuracy in thermal state estimation over a wide temperature range. We also demonstrate that the accuracy of thermal state preparation can be further enhanced by refining the reward function in the reinforcement learning process.

In scenario (2), the noiseless agent interacts with a noisy environment, where the entropy and the expectation value of the SYK Hamiltonian are computed in the presence of 1- and 2-qubit gate noise in PQC$_1$ and PQC$_2$. The qubits follow the constrained connectivity of the \texttt{IBM Eagle r3} processor. Our results indicate that for certain temperatures, the RL agent achieves better accuracy in the energy expectation value and entropy compared to scenario (1). However, when both quantities are combined to calculate the free energy, the accuracy degrades compared to the noiseless scenario. This can be tackled by utilizing sophisticated error mitigation techniques.

Finally, we evaluate the noise resilience of the parameterized circuits proposed by the RL agent by sampling the best-performing circuits from scenario (1) and running them on the \texttt{IBM Eagle r3} processor. Although the accuracy in the energy expectation value and entropy experiences a significant decrease due to noise in the QPU, its impact on the free energy is less pronounced.

The remainder of the sections is organized as follows: in Sec.~\ref{sec:prelims} we elaborate on the preliminaries, such as the problem statement and the variational thermal state preparation algorithm. This is followed by an introduction to the variational preparation of the thermal states in the RL framework in Sec.~\ref{sec:rlvqtsp}. In Sec.~\ref{sec:results} we thoroughly discuss the main results of the paper, including the efficiency of the RL framework in noiseless and realistic noisy scenarios. Finally, in Sec.~\ref{sec:conclusion} we conclude the paper and discuss possible future work following the success of this paper.

\section{Preliminaries}
\label{sec:prelims}

{The dense Sachdev-Ye-Kitaev (SYK) model describes $N$ Majorana fermions with all-to-all $q$-body interactions governed by the Hamiltonian:
}
{\begin{equation}
H = \frac{(i)^{q/2}}{q!} \sum_{1 \leq i_1 < i_2 < \cdots < i_q \leq N} J_{i_1i_2\ldots i_q} \chi_{i_1} \chi_{i_2} \cdots \chi_{i_q},
\label{eq:SYK_main} 
\end{equation}
}
{where the Majorana operators $\chi_i$ satisfy $\{\chi_i, \chi_j\} = \delta_{ij}$, and the antisymmetric random couplings $J_{i_1\ldots i_q}$ are drawn from a Gaussian distribution possessing zero mean and variance $\langle J_{i_1\ldots i_q}^2 \rangle \propto J^2/N^{q-1}$.
}
{\begin{equation}
\overline{J_{i_1i_2i_3i_4}}=0,\quad \overline{J_{i_1i_2i_3i_4}^2} = \frac{3!}{N^3} \quad (J=1).
\label{eq:Jijkl}
\end{equation}
}
{In this work, we focus on the case $q=4$, where the interactions are quartic and randomly distributed. The dimension of the Hilbert space is $\dim(\mathcal{H}) = 2^{N/2}$, where $n = N/2$ is the number of qubits. This follows from the fact that two Majorana fermions can be represented by one complex spinless fermion, which corresponds to a single qubit. While the SYK model is analytically solvable in the large-$q$ limit \cite{maldacena2016remarks}, we constrain this research to $q=4$ as provided in Eq.~\eqref{eq:SYK_main}. For simplicity, we denote $\beta J$ as $\beta$.
}
{
The SYK model's dense interaction structure presents formidable computational barriers rooted in its system-size scaling behavior. For quartic couplings ($q=4$), the Hamiltonian term count scales combinatorially as $\mathcal{O}(N^4)$, driving exponentially growing resource requirements for classical simulations. This scaling collapse motivates the pursuit of quantum advantage in simulating such correlated fermionic systems.
This scaling leads to two primary challenges (1) in classical simulation: The storage requirements grow as $\binom{N}{4}$, making exact diagonalization infeasible for large $N$. Meanwhile, in Quantum simulations, the number of gates required for parameterized quantum circuits (PQCs) increases polynomially with $N$, with a rapid growth in $\texttt{CNOT}$ gate count.
}
{To address these challenges, sparsified versions of the SYK model have been proposed in prior works \cite{garcia2021sparse, xu2020sparse}, which reduce the number of terms in the Hamiltonian while approximating its essential physics. However, even with sparsification, quantum simulations remain resource-intensive. For instance, prior work on thermal state preparation for the SYK model using variational quantum algorithms was limited to $N=12$ Majorana fermions due to these overheads \cite{araz2024thermal}.
}
{In this work, we tackle these computational bottlenecks by employing a reinforcement learning (RL) framework with an $\epsilon$-greedy policy. Our RL agent is implemented using a 3D convolutional neural network (CNN) that encodes quantum circuits as input tensors. This approach dynamically explores and optimizes circuit architectures while reducing $\texttt{CNOT}$ gate overhead by over two orders of magnitude compared to standard PQCs arising due to the real-time dynamics of SYK models. Consequently, our method enables simulations of systems with more than $N=12$ Majorana fermions while preserving model fidelity.
}

\subsection{Variational thermal state preparation~\label{sec:VTSP}}

Thermal density matrices $\rho_\beta \propto e^{-\beta H}$ play a fundamental role in characterizing equilibrium properties of quantum systems at finite temperatures. Their preparation poses greater complexity compared to limiting cases: ground states ($\beta \to \infty$, pure states) and infinite-temperature states ($\beta \to 0$, maximally mixed), which admit efficient classical representations. The intermediate regime of finite $\beta$ proves particularly crucial for studying strongly correlated systems like the Sachdev-Ye-Kitaev model, where thermalization dynamics and holographic signatures emerge through entanglement structure and out-of-time-order correlations. To prepare $\rho_\beta$, we minimize the Helmholtz free energy \(F\), given by \(F = \langle H \rangle_\beta - T S_v\),
with the temperature is given by \( 1 / \beta\), \(\langle H \rangle_\beta = \text{Tr}[\rho_\beta H]\) is the energy expectation value, where \(\rho_\beta\) is the thermal state at temperature $T$, and \({S}_v = -\text{Tr}[\rho_\beta \log \rho_\beta]\) is the von Neumann entropy. We adopt the algorithm from Ref.~\cite{selisko2023extending} whose subroutines are defined in the following. {Appendix~\ref{appendix:why_vqtsp_works} contains a comprehensive discussion of the theoretical foundations that explain why the variational thermal state preparation algorithm is effective.}

    \paragraph{{Initialization:}} Begin with all qubits initialized in the state \(\vert 0 \rangle^{\otimes n}\).
    
    \paragraph{First variational circuit PQC$_1$}\footnote{PQC$_1$ and PQC$_1$($\vec{\theta}$) are interchangeably used throughout the paper, but they carry the same meaning.}
    \begin{itemize}
        \item Ansatz construction: We utilize a hardware-efficient ansatz comprising layers of \texttt{RZ}-\texttt{RY}-\texttt{RZ} gates on each qubit, followed by CNOT entangling gates between qubits, as previously described by ref.~\cite{araz2024thermal}. This structure is chosen for its ease of implementation on real quantum hardware, offering a practical approach to circuit design.
        \item {Measurement and entropy computation:}
        First, implement the $\mathrm{PQC}_1(\boldsymbol{\theta})$ with tunable parameters $\boldsymbol{\theta} \in \mathbb{R}^d$. This prepares the variational state:
        \[
        \ket{\psi(\boldsymbol{\theta})} = \mathrm{PQC}_1(\boldsymbol{\theta})\ket{0}^{\otimes n},
        \]
        where $n$ denotes the number of qubits. Subsequent computational basis measurements yield probability distributions $\{p_i(\boldsymbol{\theta})\}_{i=1}^{2^n}$, where $p_i(\boldsymbol{\theta}) = 
        |\braket{i}{\psi(\boldsymbol{\theta})}|^2$ corresponds to the measurement outcome probability for basis state $\ket{i}$. The system's Shannon entropy is then computed as:
        \[
        S(\boldsymbol{\theta}) = -\sum_{i=1}^{2^n} p_i(\boldsymbol{\theta}) \ln p_i(\boldsymbol{\theta}),
        \]
        quantifying the information entropy of the prepared quantum state. {An elaborated discussion of the resource requirements in entropy estimation is provided in Appendix.~\ref{appndix:entropy_measurement}}
        
    \end{itemize}
    
    \paragraph{{Second variational circuit (PQC$_2$):}}\footnote{PQC$_2$ and PQC$_2$($\vec{\phi}$) are interchangeably used throughout the paper, but they carry the same meaning.}
     \begin{itemize}   
        
        \item {Initialization:}
        {After measuring PQC$_1$ in the computational basis, the state reduces to:
    \begin{equation}
    \rho_{PQC1} = \sum_i p_i(\vec{\theta}) |b_i\rangle\langle b_i|,
    \end{equation} which works as the initial state for PQC$_2$.}
    \item {Ansatz construction:}
            Employ a problem-specific or quantum hardware-inspired ansatz~\cite{herasymenko2021diagrammatic} with gates either derived from first-order Trotterization or obtained from the native gateset from quantum hardware.
        \item {Thermal state approximation:}
            Apply the unitary transformation \(\text{PQC}_2({ \vec{\phi}})\), where \({\vec{\phi}}\) are the variational parameters. Map the measured basis states to a superposition of orthonormal states. The final density matrix after PQC$_2$ is:
            \[
            \rho_{\mathrm{PQC}_2} = \sum_i |a_i({\vec{\theta}})|^2 \ket{\Psi_i({ \vec{\phi}})}\bra{\Psi_i({\vec{\phi}})},
            \]
            where \(\ket{\Psi_i({\vec{\phi}})} = \text{PQC}_2({ \vec{\phi}})\ket{b_i}\).
    
        \item {Free energy minimization:} Minimize the free energy \(F = \langle H \rangle_\beta - T S\) using a classical optimizer to refine the parameters \({ \theta}\) and \({ \phi}\).
\end{itemize}

\paragraph{{Performance evaluation:}}
        Evaluate the free energy, entropy, and energy expectation value and compare with theoretical values.

\section{Variational thermal state preparation with reinforcement learning}\label{sec:rlvqtsp}

We give an overview of the variational thermal state preparation of the SYK model within the reinforcement learning (RL) framework, for details of the framework see Appendix~\ref{appndix:rl_framework}. The algorithm is illustrated in Fig.~\ref{fig:main_algorithm}. The framework is used to estimate the parameters of the PQC$_1$ and the structure as well as parameters of the PQC$_2$ elaborated in Sec.~\ref{sec:VTSP}. Wherein we present state, the action representations, and the reward function used in this work.

\paragraph{The state} In the hybrid quantum-classical environment, the agent begins each episode with PQC$_1$ (which includes a layer of \texttt{RZ}-\texttt{RY}-\texttt{RZ} rotations on each qubit followed by CNOT entangling gate connected in {cyclic manner}), then it sequentially adds gates to the PQC$_2$ (which starts as an empty circuit at the beginning of each episode) until the maximum number of actions is reached. At each step, the state is represented by the current circuit, i.e., PQC$_1$ combined with PQC$_2$, and the action is to append a gate. To enable deep RL methods, states and actions are encoded in a format suitable for neural networks. Specifically, each state is represented by a tensor-based binary encoding of the circuit’s gate structure, as described in Appendix~\ref{appndix:tensor_encoding}. The action space consists of CNOT gates and 1-qubit parameterized rotation gates (\texttt{RX}, \texttt{RY}, and \texttt{RZ}). Notably, we omit the continuous parameters of the rotation gates, focusing instead on the estimated energy of the circuit to define the state representation for the RL agent.

\paragraph{The agent} The reinforcement learning (RL) agent utilizes a 3D convolutional neural network (CNN), as the quantum circuit is encoded as a 3D tensor (refer to Appendix~\ref{appndix:tensor_encoding}). In prior studies~\cite{kundu2024enhancing,kundu2024kanqas,patel2024curriculum,fosel2021quantum}, this 3D representation was typically flattened into a 1D format to be processed by a fully connected feedforward neural network (FNN)~\cite{sazli2006brief,svozil1997introduction}. However, flattening 3D data into 1D can result in the loss of critical spatial information, diminished model expressiveness, and potentially suboptimal performance. Architectures like 3D CNNs that can directly process the 3D structure are often preferable when working with volumetric data. If not flattened when operating on a structure like a 3D-CNN~\cite{jin2014flattened,rao2019natural}, the neural net receives a dataset with distinguishable spatial information.
In the case of the 3D binary encoding, where the X-axis represents the position of gates in PQC$_2$, the Y-axis denotes the type of gate, and the Z-direction is the depth of PQC$_2$. In Appendix.~\ref{appndix:CNN_vs_FNN} we elaborately show that directly feeding higher-dimensional data through a 3D-CNN provides improved training, leading to lower approximation error with more compact parameterized quantum circuits compared to processing flattened data through an FNN. The hypermeters settings of the agent is discussed in Appendix~\ref{appndix:agent_hyperparameters}.

\paragraph{The reward function} To steer the agent towards the target, we use the same reward function $r$ at every time step t of an episode, as in~\cite{ostaszewski2021reinforcement}. The reward function is $r$ defined as
{
\begin{equation}
r = 
\begin{cases} 
5, & \text{if } F_t(\beta=\text{const.},\vec{\theta},\vec{\phi}) \leq \zeta_F, \\
-5, & \text{if } F_t(\beta=\text{const.},\vec{\theta},\vec{\phi}) > \zeta_F \;\text{and,}\; \text{step no.} = D_\text{max}, \\
E_\text{term}, & \text{otherwise}.
\end{cases}
\label{case:error_rwd}
\end{equation}
}
Where $F_t(\beta=\text{const.},\vec{\theta},\vec{\phi})$ is the Helmholtz free energy at a constant temperature, at each step $t$, and $E_\text{term} = \text{max}\left(\frac{F_\text{prev} - F_\text{current}}{\lvert F_\text{prev} - F_\text{true} \rvert}, -1, 1\right)$. The $\zeta_F$ is a user-defined threshold value on the error in estimating the free energy, and $D_\text{max}$ denotes the total number of steps allowed for an episode, which can also be understood as the
maximum number of actions allowed per episode. Note that the extreme reward values $\pm$5 signal the end of an episode, leading to two stopping conditions: exceeding the threshold or reaching
the maximum number of actions. For our task, we set the $\zeta_\text{F}$ to $10^{-2}$.

In Sec.~\ref{sec:results} we show that using the reward in Eq.~\ref{case:error_rwd} we can achieve a very good approximation of the free energy, but it cannot provide as good an approximation to the energy expectation of the SYK Hamiltonian and the entropy. Hence, we engineer the reward signal to incorporate the accuracy of the prepared thermal state by calculating its fidelity from the target thermal state after PQC$_2$. Hence, the modified reward function is given by
\begin{equation}
r = 
\begin{cases} 
5, & \text{if } F_t(\beta=\text{const.},\vec{\theta},\vec{\phi}) {\leq} \zeta_F\; \text{and,}\;\textrm{Fid}(\rho({\vec{\phi}})) \geq \zeta_\textrm{Fid.} \\
-5, & \text{if } \textrm{Fid}(\rho({\vec{\phi}})) < \zeta_\textrm{Fid} \;\text{and,}\; \text{step no.} = D_\text{max}, \\
0.6 \times E_\text{term} + 0.4 \times \textrm{Fid}_\text{term}, & \text{otherwise},
\end{cases}
\label{case:error_and_fidelity_rwd}
\end{equation}
Where $\textrm{Fid}_\text{term} = 2\times\textrm{Fid}(\rho({\vec{\phi}}))-1$, $\textrm{Fid}(\rho({\vec{\phi}})) = {\mbox{Tr}} \Big(\sqrt{\rho_{\beta}^{1/2} \rho(\vec{\phi})~ \rho_{\beta}^{1/2}}\Big)$ and \(\rho_\beta\) represents the density matrix that corresponds to the true thermal state. $\zeta_\textrm{Fid}$ is the user-defined threshold on fidelity, throughout the paper, we choose $\zeta_\textrm{Fid}=0.9$. {An elaborated analysis of the reward function defined in Eq.~\ref{case:error_and_fidelity_rwd} is provided in the Appendix~\ref{appndix:rwd_function_analysis}}.

In the upcoming section, we will see that compared to the reward function Eq.~\ref{case:error_rwd} and Eq.~\ref{case:error_and_fidelity_rwd} to the results obtained in ref.~\cite{araz2024thermal}, in the Sec.~\ref{sec:results} we achieve higher accuracy in the free energy, the entropy, and the energy of the thermal state.

\section{Results}\label{sec:results}

We implement a reinforcement learning (RL)-driven hybrid quantum-classical thermal state preparation algorithm (see Ref.~\cite{selisko2023extending} for the details on the variational algorithm). The entropy-estimation circuit $\mathrm{PQC}_1$ employs entangling layers with cyclic connectivity, while $\mathrm{PQC}_2$, which calculates the expectation value, is built using the RL-agent. Numerical simulations leverage the \textsc{Qulacs} quantum simulator \cite{suzuki2021qulacs} in a heterogeneous architecture\footnote{Hybrid computation: 64-core AMD EPYC CPUs and NVIDIA A100 GPUs via CUDA-quantum integration \cite{amd2025heterogeneous}}.

Throughout the paper we use the gradient-free optimizer COBYLA~\cite{powell2007view} with $10^3$ iterations to optimize the parameters of the PQC$_1$ and the PQC$_2$. The RL framework is running for a maximum of $5\times10^3$ episodes the RL environment interacts with the agent for a predefined number of steps. Furthermore, the agent is optimized by the ADAM optimzier~\cite{kingma2014adam} with $10^{-3}$ learning rate. Further details of the RL-agent can be found in Appendix~\ref{appndix:CNN_vs_FNN} and the environment that contains the tensor-based state and the reward function is elaborated in Section~\ref{sec:rlvqtsp}.

\paragraph{Sampling the best-performing circuits} 
As discussed earlier, during the training of $5 \times 10^3$ episodes, the RL agent generates a class of parameterized quantum circuits with varying depths and gate counts, meeting predefined thresholds for free energy and fidelity. Selecting a specific circuit that accurately approximates the free energy, Hamiltonian energy expectation, and entropy is a challenging task. To streamline this process, we employ a filtering subroutine to identify the best-performing circuit. This filter applies a threshold on the errors, defined as:  
\begin{equation}
    \varepsilon = \Delta F + w_a \Delta \langle H\rangle_\beta + w_b \Delta S, \label{eq:filter_threshold}
\end{equation}  
where $\Delta F$ represents the error in free energy, $\Delta \langle H\rangle_\beta$, scaled by the weight factor $w_a$, corresponds to the error in the Hamiltonian expectation, and $\Delta S$, scaled by $w_b$, denotes the error in entropy. Among the circuits proposed by the RL agent, the one minimizing $\varepsilon$ is selected as the best-performing candidate.

The parameters $w_a$ and $w_b$ are dependent on the reward function and the number of qubits. The parameter values used in the subsequent simulations are summarized below:  
\begin{enumerate}  
    \item \textbf{Free energy reward (Eq.~\ref{case:error_rwd}):} For $N = 8, 10,$ and $12$, we use $w_b = 0$ and $w_a = 0.5, 1.02,$ and $2$, respectively. For $N = 14$, the best-performing circuit is obtained with $w_a = 0$ and $w_b = 2$.  
    \item \textbf{Free energy and fidelity reward (Eq.~\ref{case:error_and_fidelity_rwd}):} For $N = 8, 10,$ and $12$, we use $w_b = 0$ and $w_a = 0.8, 1.02,$ and $1.16$, respectively. Similarly, for $N = 14$, the optimal circuit is found with $w_a = 0$ and $w_b = 2$.  
\end{enumerate}

{The objective of optimizing thermal state preparation by adjusting the entropy threshold $ w_b $ for the range $ 8 < N < 14 $ arises from our analysis of small Majorana fermionic systems. In these systems, the various quantum circuit structures proposed by the RL-agent provide a good approximation to the entropy of the thermal state due to the relatively small system size. This contrasts with larger systems, where a trade-off in accuracy arises between free energy and Hamiltonian expectations.
}

{After gathering results from the RL agent, we find that for $ N = 10 $ and $ N = 12 $, it is important to slightly increase the weight associated with the filter of $ \Delta \langle H \rangle_\beta $. Since the primary goal of the RL agent is to optimize free energy, this adjustment helps improve the accuracy of the thermal state representation. However, for $ N = 14 $, while the RL agent achieves a good approximation of free energy, we encounter limitations in entropy accuracy that become significant bottlenecks. To address this issue, we propose switching off the weight of $ \Delta \langle H \rangle_\beta $ and instead applying equal weight to the entropy filter. This approach not only enhances the approximation of free energy but also reduces errors in both entropy and $ \Delta \langle H \rangle_\beta $. 
}

{Overall, this post-processing step is crucial for refining our results and ensuring a more accurate representation of thermal states. In Appendix~\ref{appndix:art_of_sampling}, we illustrate the impact of a wide range of weights on the sampling of the best thermal state from all possible solutions proposed by the RL-agent.
}
\subsection{Noiseless simulation}
\begin{figure}[hbt!]
\centering
\begin{subfigure}{.48\linewidth}
  \includegraphics[width=\linewidth]{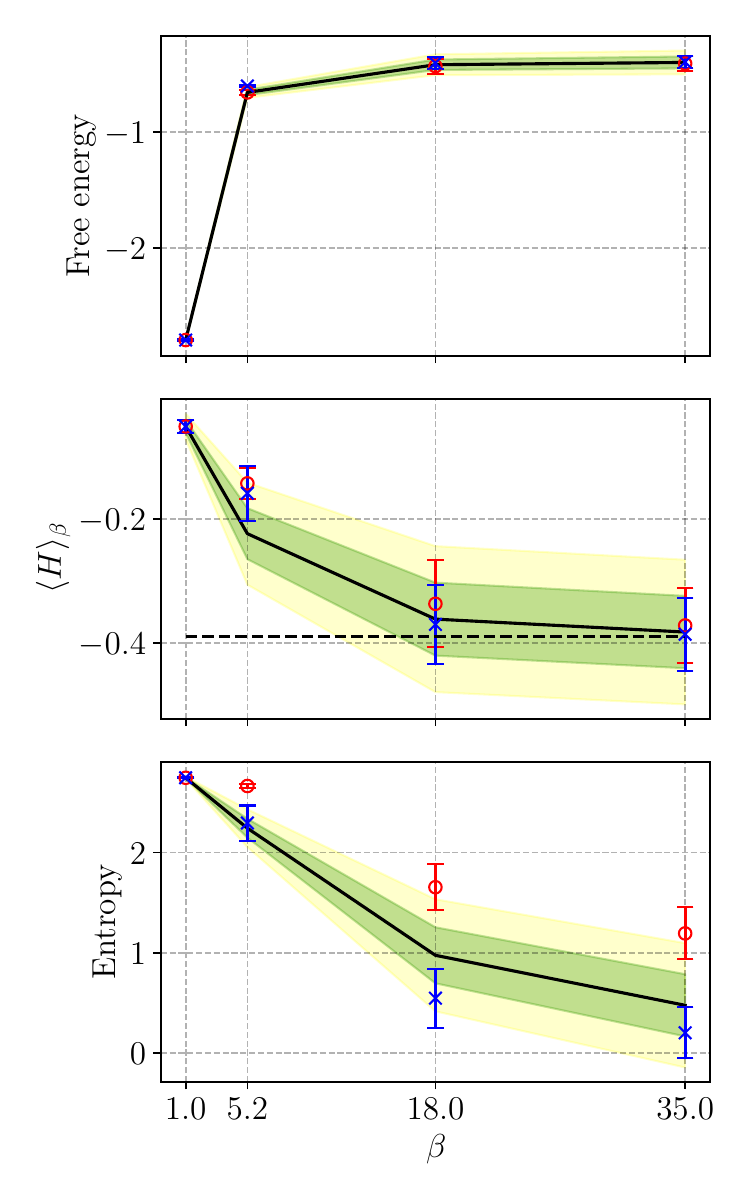}
  \caption{N=8}
  \label{MLEDdet}
\end{subfigure}
\begin{subfigure}{.48\linewidth}
  \includegraphics[width=\linewidth]{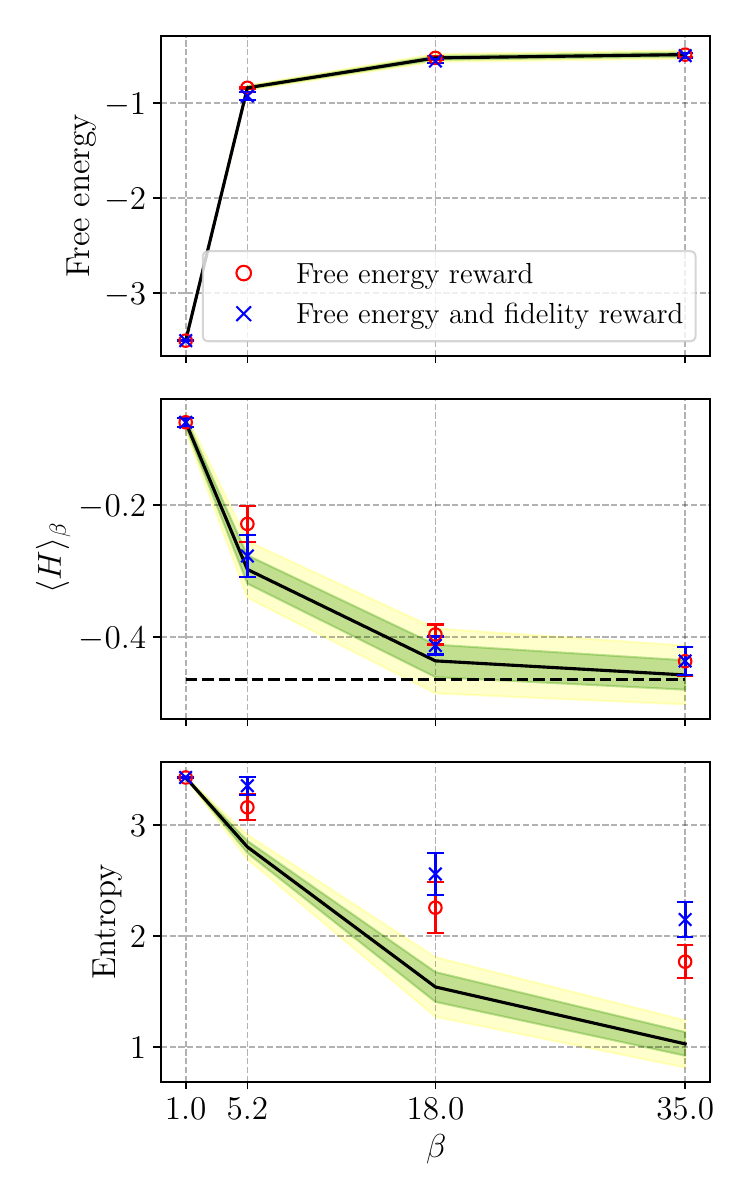}
  \caption{N=10}
  \label{energydetPSK}
\end{subfigure}
\caption{\small \textbf{RL framework enabling ground state preparation for $N =8,10$ Majorana fermions with high fidelity $F\geq0.8$.} The number of gates required to achieve the result is at least 10-fold lower than the first-order Trotterized Hamiltonian-inspired ansatz for smaller systems. Red dots show RL-agent performance using free energy error as reward (Eq.~\ref{case:error_rwd}). Blue crosses depict RL circuit performance using both free energy error and thermal state fidelity as rewards (Eq.~\ref{case:error_and_fidelity_rwd}). The solid line represents the exact mean, with green (yellow) regions indicating $1\sigma$ ($2\sigma$) deviations.}
\label{fig:noiseless_training1}
\end{figure}
\begin{figure}[hbt!]
\centering
\begin{subfigure}{.48\linewidth}
  \includegraphics[width=\linewidth]{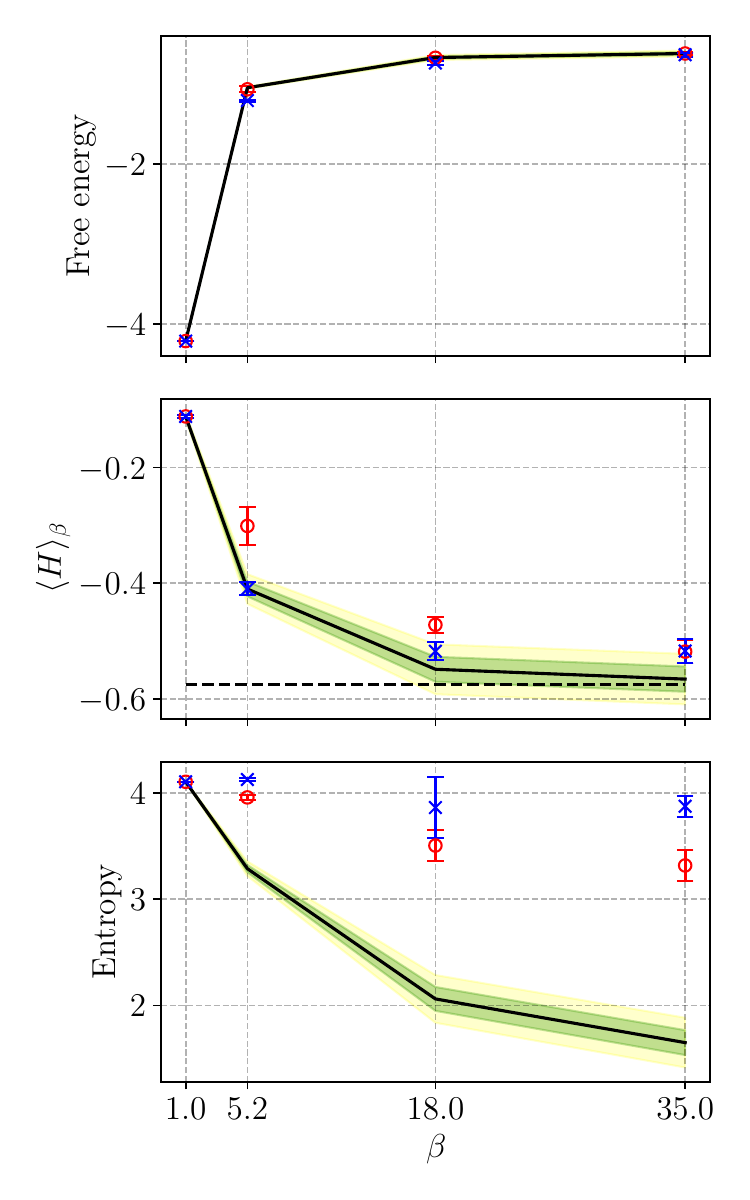}
  \caption{N=12}
  \label{velcomp}
\end{subfigure} 
\begin{subfigure}{.48\linewidth}
  \includegraphics[width=\linewidth]{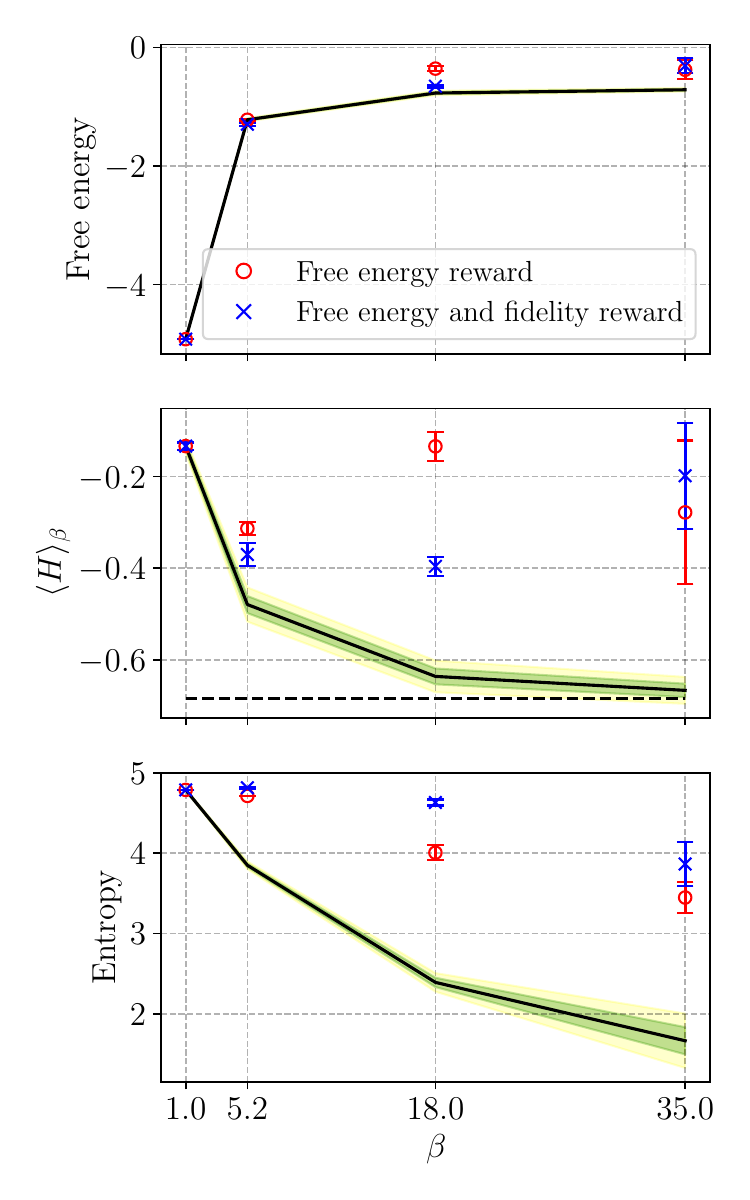}
  \caption{N=14}
  \label{estcomp}
\end{subfigure}
\caption{\small \textbf{RL framework surpasses the bottleneck in ref.~\cite{araz2024thermal}, enabling ground state preparation for $N > 12$ Majorana fermions.} It achieves efficient gate scaling, facilitating thermal state preparation for $N \leq 14$ with fidelity $F \geq 0.8$ for $N = 12$. As stated in Fig.~\ref{fig:noiseless_training1}, Red dots show RL-agent performance using free energy error as reward (Eq.~\ref{case:error_rwd}). Blue crosses depict RL circuit performance using both free energy error and thermal state fidelity as rewards (Eq.~\ref{case:error_and_fidelity_rwd}). The solid line represents the exact mean, with green (yellow) regions indicating $1\sigma$ ($2\sigma$) deviations.}
\label{fig:noiseless_training2}
\end{figure}

Figures~\ref{fig:noiseless_training1} and \ref{fig:noiseless_training2} demonstrate our framework's performance for the dense SYK model across system sizes $N=8,10$ (Fig.~\ref{fig:noiseless_training1}) and $N=14$ (Fig.~\ref{fig:noiseless_training2}). Five disorder realizations were simulated per system size, with agent retraining per instance to ensure robust benchmarking. Each panel quantifies three thermodynamic quantities: the free energy $F = -\beta^{-1}\ln Z$, energy expectation $\langle H \rangle$, and entropy $S = \beta(\langle H \rangle - F)$.
Shaded regions (green: $1\sigma$, yellow: $2\sigma$) represent statistical deviations from exact diagonalization results (solid lines). To this end, two reward strategies are compared:
\begin{itemize}
    \item \textit{Red markers}: Free energy minimization via reward.~\ref{case:error_rwd}.
    \item \textit{Blue markers}: Combined free energy and state fidelity rewards in Eq.~\ref{case:error_and_fidelity_rwd}.
\end{itemize}
Error bars reflect standard deviations across disorder realizations. The fidelity $F(\rho(\boldsymbol{\phi}))$ remains high ($0.80$–$0.98$) for $8 \leq N \leq 12$, but degrades substantially ($0.4$–$0.80$) at $N=14$, highlighting the challenge of maintaining state quality in larger systems.

In the RL framework, the scaling of the number of gates required is substantially lower compared to the scaling in first-order Trotterization. This efficient scaling enables the preparation of thermal states for increasing numbers of Majorana fermions using a minimal number of 1- and 2-qubit gates. Consequently, this study achieves, for the first time, a variational preparation of thermal states for $N=14$ Majorana fermions, demonstrating good agreement in estimating the free energy. Notably, at $\beta=5.2$, the RL agent provided the most accurate approximation of the free energy and expectation value compared to other temperatures.

The noiseless simulation results exhibit very good agreement in the free energy, Hamiltonian expectation, and entropy for $N=8$. However, as the number of Majorana fermions increases, the accuracy of the energy expectation and fidelity diminishes. Specifically, the error in estimating the Shannon entropy is significantly higher than the error in the Hamiltonian expectation value, although the free energy estimation error remains low.

Additionally, we observe that incorporating fidelity of the thermal state into the reward signal (blue crosses in Fig.~\ref{fig:noiseless_training1} and Fig.~\ref{fig:noiseless_training2}) improves the accuracy of the energy expectation value. This improvement stems from the fidelity being evaluated alongside the expectation value immediately after executing PQC$_2$. Enhanced fidelity in thermal state preparation directly improves the construction of PQC$_2$, thereby refining the accuracy of the Hamiltonian expectation value estimation.

In summary, the results highlight the effectiveness of the RL framework in variationally preparing thermal states when the system size scales. By incorporating fidelity into the reward mechanism, the RL approach achieves enhanced accuracy in energy expectation estimation, demonstrating its robustness and scalability. These findings encourage the exploration of the RL-driven quantum state preparation in noisy and hardware-contained scenarios.

\begin{figure}[h!]
    \centering
    \includegraphics[width=\linewidth]{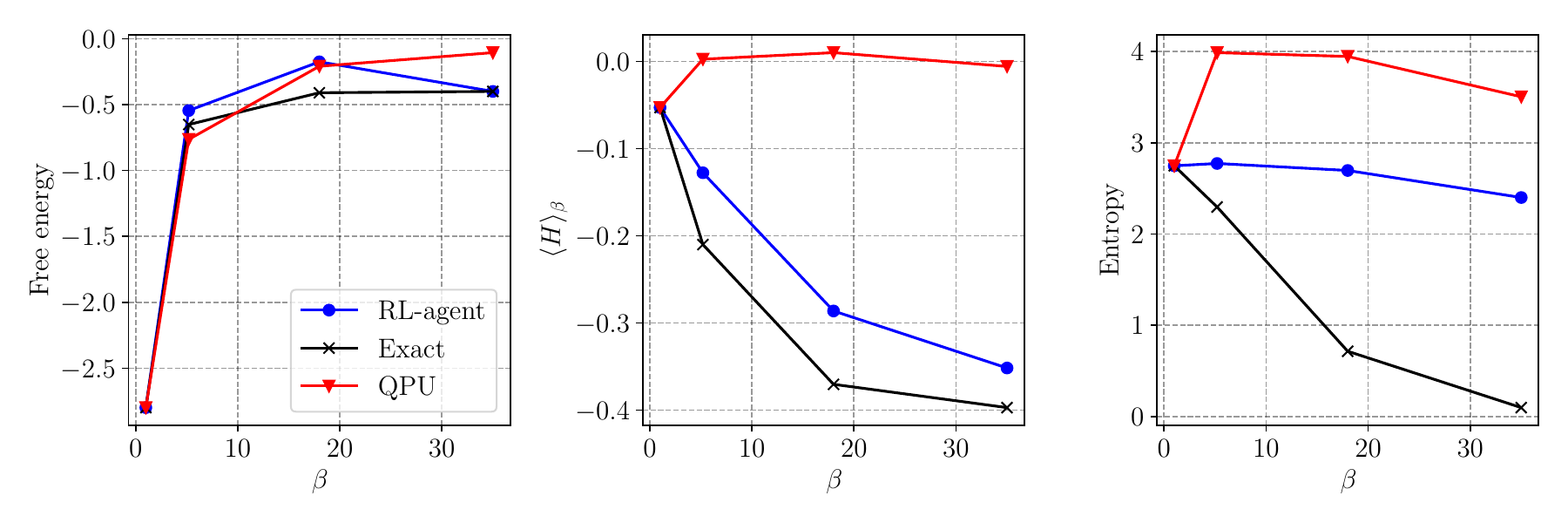}
    \caption{\small\textbf{The RL-agent selected circuit excels in approximating free energy for $N=8$, even with hardware imperfections}. QPU noise significantly impacts Hamiltonian expectation value and entropy, yet free energy remains largely unaffected, demonstrating the RL-agent's effectiveness in both ideal and noisy environments. This robustness suggests potential for training in realistic noisy conditions. No error-mitigation techniques were applied to the QPU.}
    \label{fig:noise_circ_onqpu}
\end{figure}
\subsection{Noise robustness on QPU}

In this section, we study the robustness of the circuits selected by the RL-agent under realistic noisy conditions. The RL-agent is trained in a noiseless environment to optimize the parameterized circuits for minimizing the free energy. After training, the best circuit is selected according to how accurately it can find the free energy and then the circuit is executed on the \texttt{IBM Eagle r3} processor to assess its practical performance. In the absence of any sophisticated error-mitigation techniques. See Appendix~\ref{appndix:QPU} for further information on the quantum hardware noise and limitations. The results were compared with exact classical calculations and experimental outcomes on the QPU for different temperatures. 
\begin{table}[ht]
\caption{\small\textbf{Number of gates after transpilation on \texttt{IBM Eagle r3} processor}.}
\vspace{7pt}
\centering
\begin{tabular}{|c|c|c|c|c|}
\hline
To optimize & Gate type & $\beta=5.2$ & $\beta=18$ & $\beta=35$ \\ \hline
\multirow{2}{4em}{Entropy} & 1-qubit & 63 & 69 & 59 \\
 &2-qubit & 7 & 9 & 10 \\ \hline
 \multirow{2}{5em}{Expectation value} & 1-qubit & 72 & 79 & 71 \\
 &2-qubit & 9 & 10 & 10 \\ \hline
\end{tabular}
\label{tab:qpu_transpile}
\end{table}

Fig.~\ref{fig:noise_circ_onqpu} depicts the free energy, expectation value of the Hamiltonian, and entropy as functions of the inverse temperature $\beta$. The RL-selected circuit demonstrates good agreement with the exact theoretical results in the noiseless scenario. When deployed on the QPU, the error in the Hamiltonian expectation value and the entropy increase dramatically due to hardware imperfections, but the error in the free energy stays relatively lower and comparable to the noiseless scenario. This is because in this run the RL-agent is specifically trained to give a parameterized circuit that provides a proper approximation to the free energy, albeit the errors in the energy expectation value and the entropy are high. The number of gates after transpilation in the QPU is provided in Tab.~\ref{tab:qpu_transpile}.

These results highlight the effectiveness of the RL-agent in identifying circuits that are not only optimal in idealized settings but also, to some extent, robust to QPU. The trade-off between accuracy and hardware limitations is evident, particularly at higher values of $\beta$, where noise effects become more pronounced. This tradeoff motivated us to train the agent in a realistic noisy environment imposed with the qubit connectivity constraints in the upcoming section. 

\begin{figure}[h!]
    \centering
    \includegraphics[width=\linewidth]{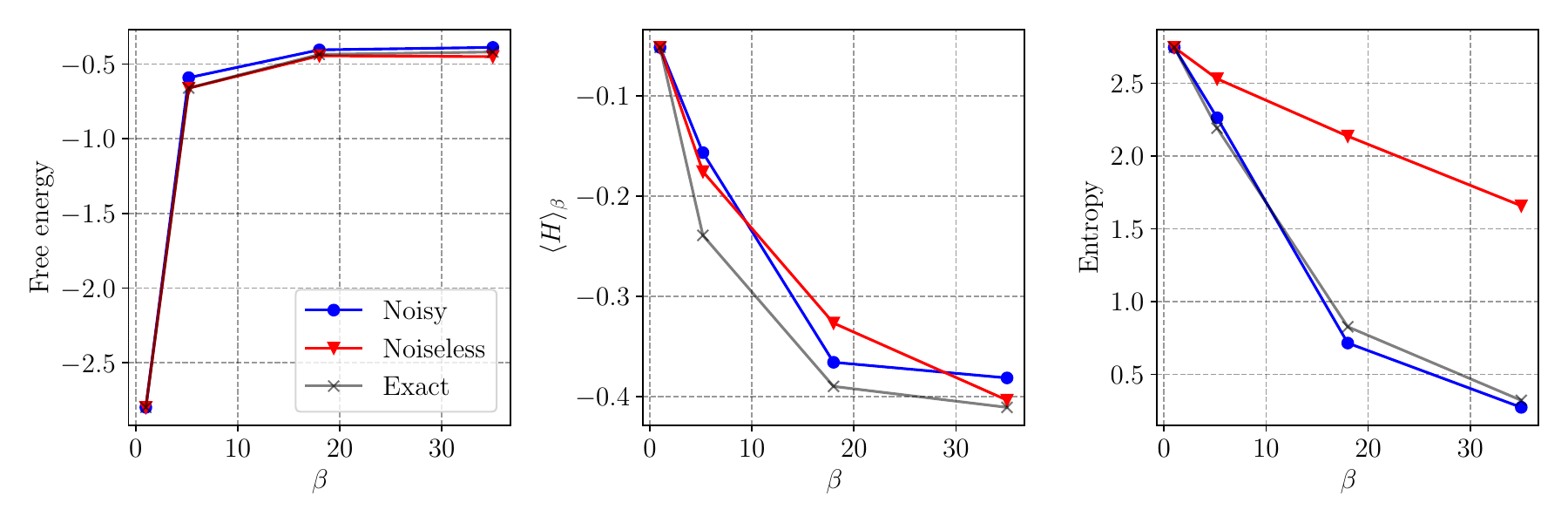}
    \caption{\small\textbf{Impact of noise and hardware connectivity constraints on thermal state preparation in the RL framework for $N=8$ Majorana fermions}. The free energy accuracy decreases under realistic noise conditions (bit-flip and depolarizing noise), while the accuracy of the energy expectation value and entropy improves, showcasing the trade-offs introduced by noise in hardware-constrained quantum simulations.}
    \label{fig:noisy_simulation}
\end{figure}

\subsection{Noisy training with constrained qubit connectivity}

In this section, we consider the variational framework is noisy, i.e., each gate in the action space is subjected to quantum noise, leading to a noisy PQC$_1$ and PQC$_2$ where the noise values correspond to the median noise of \texttt{IBM\_brisbane} device. As the action space consists of 1- and 2-qubit gates, we add bit flip noise of strength $2.342\times10^{-4}$ after each application of a 1-qubit gate, and 2-qubit depolarizing noise of strength $8.043\times10^{-3}$ is applied after each CNOT gate. Hence the RL agent interacts with a noisy environment and receives a noisy reward signal.

The results presented in Fig.~\ref{fig:noisy_simulation} highlight the impact of noise during the simulation for $N=8$ Majorana fermions, executed under the constrained connectivity of the \texttt{IBM Eagle r3} processor. A key observation is the improved accuracy of the energy expectation value and the entropy in the noisy scenario compared to the noiseless simulations. This improvement may be attributed to the adaptive optimization of the RL agent, which dynamically accounts for noise during the construction of the parameterized quantum circuit.

However, the cumulative accuracy of the free energy estimation diminishes noticeably in the presence of noise. This reduction is due to the cumulative effects of noise during the execution of the circuits. Unlike the energy expectation value and entropy, which are influenced by local noise-induced fluctuations, the free energy calculation depends on a more global accuracy of the thermal state, making it more susceptible to cumulative errors introduced by gate noise and connectivity constraints.

Hence, this leads us to conclude that the adaptive capabilities of the RL-agent allow for improved performance in certain metrics, such as energy expectation and entropy; the free energy is a more challenging quantity to optimize under realistic noise models. This analysis emphasizes the need for further enhancements in noise mitigation strategies~\cite{guo2022quantum,filippov2023scalable,botelho2022error,robbiati2023real,cai2023quantum}.

\subsection{Improvement in CNOT count}

This section presents the improvement in the number of CNOT gates in the RL-agent proposed circuit compared to first-order Trotterization. The improvement is quantified by
\begin{equation}
\text{Improvement} = \frac{\text{CNOTs}_{\text{trotterized}}}{\text{CNOTs}_{\text{RL-agent}}},
\end{equation}

for different qubit counts and values of $\beta$. In Tab.~\ref{tab:improvement_cnot} the results show that for 5 and 6 qubits, the RL-agent offers over a $10^2$-fold improvement in the number of CNOT gates compared to first-order Trotterization-based Hamiltonian variational ansatz~\cite{araz2024thermal}, particularly for higher values of $\beta$. This demonstrates the ability of the RL-agent to significantly reduce circuit complexity as the qubit count increases, making it favorable for scaling to higher qubits.
\begin{table}[ht]
\caption{\small\textbf{Improvement in the number of CNOT gates in the RL-agent proposed circuit compared to first-order Trotterization}. The improvement is quantified as the ratio of the number of CNOT gates in the trotterized parameterized circuit to that in the RL-agent proposed circuit. On the \textbf{left}, we calculate the improvement using the free energy reward signal (as defined in Eq.\ref{case:error_rwd}), while on the \textbf{right}, we consider the improvement with both free energy and fidelity as reward signals (as defined in Eq.\ref{case:error_and_fidelity_rwd}).}
\vspace{7pt}
\centering
\begin{tabular}{|c|c|c|c|}
\hline
$\beta$ & $N=8$ & $N=10$ & $N=12$ \\ \hline
$5.2$  & 32.7957 & 46.0322 & 242.7860 \\ \hline
$18$   & 27.7726 & 54.7286 & 189.2048 \\ \hline
$35$   & 84.6780 & 85.7093 & 349.0837 \\ \hline
\end{tabular}
\hspace{1cm}
\begin{tabular}{|c|c|c|c|}
\hline
$\beta$ & $N=8$ & $N=10$ & $N=12$ \\ \hline
$5.2$  & 46.1245 & 119.2494 & 571.6576 \\ \hline
$18$   & 57.2831 & 68.1234  & 468.7002 \\ \hline
$35$   & 15.5273 & 41.5346  & 406.7558 \\ \hline
\end{tabular}
\label{tab:improvement_cnot}
\end{table}

\section{Conclusion and future work}\label{sec:conclusion}
\paragraph{Summary} In this work, we present a novel approach to preparing thermal states of the Sachdev-Ye-Kitaev (SYK) model using a variational quantum algorithm in the reinforcement learning (RL) framework. In the variational algorithm, which plays the role of the environment in the RL framework, the parameterized quantum circuits are encoded in a 3D tensor. In the encoding, the X-axis represents the position of gates, the Y-axis denotes the type of gate, and the Z-direction is the depth. To process such higher-dimensional data, we integrate the environment with a 3D convolutional neural network (CNN) to optimize the preparation of the thermal states for different temperatures.

The 3D-CNN is guided by a $\epsilon$-greedy policy, and a composite reward signal helps the agent to choose an action (i.e., a quantum gate) for a specific state of the environment. In this framework we operate in two key scenarios:
(1) A fully noiseless environment for systems with up to $N = 14$ Majorana fermions. (2) A noisy environment with constrained qubit connectivity as per the \texttt{IBM Eagle r3} processor for $N = 8$ Majorana fermions, using a noiseless agent.

This RL framework has proven highly effective in identifying optimal parameterized quantum circuits for thermal state preparation across a wide range of temperatures. Crucially, our approach significantly reduces the computational overhead associated with this task, making it feasible to prepare thermal states for larger SYK systems, $N\geq14$, which was previously not feasible. Our method significantly reduces the computational overhead associated with preparing thermal states for large SYK systems ($N > 12$ Majorana fermions) on near-term quantum devices. Key findings of the paper include:

\begin{enumerate}
    \item A two-order-of-magnitude reduction in the number of CNOT gates for systems $N \geq 12$ compared to traditional methods like first-order Trotterization.
    
    \item Effective thermal state preparation in both noiseless and noisy quantum hardware environments, maintaining high accuracy.

    \item Successful integration of reinforcement learning with convolutional neural networks to iteratively refine quantum circuits and their parameters.
\end{enumerate}

{While our current implementation does not match the system sizes achievable in classical simulations of the SYK model ({up to $N\sim60$~\cite{kobrin2021many}}), this work demonstrates important progress in RL-optimized quantum circuits. The methodology developed here has broader implications for quantum simulation, particularly for studying quantum gravity phenomena on near-term hardware.}

\paragraph{Future work}

Building on the success of this study, several promising directions for future research emerge:

\begin{enumerate}
    \item {\textbf{Scaling to larger systems by gadgets:} We propose to investigate the scalability of our approach for SYK models with $N > 14$ Majorana fermions, pushing the boundaries of current quantum simulations. This can be achieved by leveraging the recently proposed gadget reinforcement learning \cite{kundu2024easy}, which enables the learning of composite gate sets from smaller Majorana fermionic SYK models. These learned gate sets can then be efficiently applied to solve models with a larger number of Majorana fermions, enhancing scalability and providing insights into the structure of the quantum circuits, including identifying optimal gate sequences, circuit compression, and potential generalization to other quantum systems or problems.}

    \item \textbf{Out-of-time-order correlations:} Apply this method to accurately compute out-of-time-order correlations (OTOCs) for the SYK model~\cite{asaduzzaman2023model}, which are crucial for studying quantum chaos and holographic properties.

    \item \textbf{Exploring reward engineering:} {Explore other forms of reward function to benchmark the performance of the CNN agent. Such as, we can encode the information corresponding to the properties of the thermal state (e.g., the entropy, purity, etc.) to improve the thermal state preparation further. For instance, extending Eq.~\ref{case:error_and_fidelity_rwd} to include weighted entropy optimization:
    \begin{equation}
        r = 
        \begin{cases} 
        5, & \text{if } F_t \leq \zeta_F \land \textrm{Fid}(\rho({\vec{\phi}})) \geq \zeta_\textrm{Fid} \land \textrm{Ent.}_\text{term} \leq \zeta_\text{Ent.}, \\
        -5, & \text{if } \textrm{Fid}(\rho({\vec{\phi}})) < \zeta_\textrm{Fid} \land \text{step no.} = D_\text{max}, \\
        a E_\text{term} + b \textrm{Fid}_\text{term} + c \textrm{Ent.}_\text{term}, & \text{otherwise},
        \end{cases}
    \end{equation}
    where $c$ controls entropy weighting, could better align state preparation with specific observables.}

    \item {\textbf{Transfer learning and policy generalization:} In this study, the RL-agent is trained from scratch for each inverse temperature $\beta$ and for different values of $N$. However, one could implement transfer learning strategies~\cite{tan2018survey} to adapt the policy learned for a specific $N$ and reuse it across different inverse temperatures.}

    \item \textbf{Quantum information-theoretic analysis:} Recent quantum informatic research suggests analyzing RL-agent-produced circuits can significantly enhance agent performance~\cite{sadhu2024quantum}. By applying quantum information-theoretic principles to circuit analysis, we can optimize circuit architectures, identify performance bottlenecks, and refine the RL framework's efficiency.
    
    \item \textbf{Advanced error mitigation:} Implement sophisticated error mitigation techniques such as~\cite{cai2021multi, filippov2023scalable,cai2023quantum,endo2018practical,botelho2022error} to improve the accuracy of results in noisy quantum environments, particularly for free energy calculations for larger $N$.

    \item \textbf{Generalization to other models:} Extend the RL framework to optimize quantum circuits for other strongly correlated quantum systems beyond the SYK model.

\end{enumerate}

\section{Acknowledgment} The author expresses my gratitude to Sara Doelman for her insightful discussions on convolutional neural networks and her encouragement to publish these results. The author also thanks Raghav Govind Jha for the conversations that inspired this paper. Additionally, the author is grateful to Guilherme Ilário Correr, Aritra Sarkar, Abhishek Sadhu, Jarosław Adam Miszczak, and Stefano Mangini for their valuable suggestions and comments that improved the quality of this article. Finally, The author wishes to acknowledge \textit{CSC–IT Center for Science, Finland}, for computational resources. The author acknowledges funding from the Research Council of Finland through the Finnish Quantum Flagship project 358878 (UH).

\bibliographystyle{unsrt}
\bibliography{ref}

\begin{thebibliography}{10}

\bibitem{witten1998anti}
Edward Witten.
\newblock Anti de sitter space and holography.
\newblock {\em arXiv preprint hep-th/9802150}, 1998.

\bibitem{sachdev1993gapless}
Subir Sachdev and Jinwu Ye.
\newblock Gapless spin-fluid ground state in a random quantum {Heisenberg} magnet.
\newblock {\em Physical review letters}, 70(21):3339, 1993.

\bibitem{kitaev2015simple}
Alexei Kitaev.
\newblock A simple model of quantum holography (part 1).
\newblock {\em Entanglement in strongly-correlated quantum matter}, page~38, 2015.

\bibitem{kitaev2015simplep2}
Alexei Kitaev.
\newblock A simple model of quantum holography (part 2).
\newblock {\em Entanglement in strongly-correlated quantum matter}, page~38, 2015.

\bibitem{maldacena2016bound}
Juan Maldacena, Stephen~H Shenker, and Douglas Stanford.
\newblock A bound on chaos.
\newblock {\em Journal of High Energy Physics}, 2016(8):1--17, 2016.

\bibitem{kim2021dirac}
Jaewon Kim, Ehud Altman, and Xiangyu Cao.
\newblock Dirac fast scramblers.
\newblock {\em Physical Review B}, 103(8):L081113, 2021.

\bibitem{brown2017quantum}
Adam~R Brown, Leonard Susskind, and Ying Zhao.
\newblock Quantum complexity and negative curvature.
\newblock {\em Physical Review D}, 95(4):045010, 2017.

\bibitem{gur2018does}
Guy Gur-Ari, Raghu Mahajan, and Abolhassan Vaezi.
\newblock Does the {SYK} model have a spin glass phase?
\newblock {\em Journal of High Energy Physics}, 2018(11):1--20, 2018.

\bibitem{kobrin2021many}
Bryce Kobrin, Zhenbin Yang, Gregory~D Kahanamoku-Meyer, Christopher~T Olund, Joel~E Moore, Douglas Stanford, and Norman~Y Yao.
\newblock Many-body chaos in the {Sachdev-Ye-Kitaev} model.
\newblock {\em Physical review letters}, 126(3):030602, 2021.

\bibitem{maldacena2016remarks}
Juan Maldacena and Douglas Stanford.
\newblock Remarks on the {Sachdev-Ye-Kitaev} model.
\newblock {\em Physical Review D}, 94(10):106002, 2016.

\bibitem{polchinski2016spectrum}
Joseph Polchinski and Vladimir Rosenhaus.
\newblock The spectrum in the {Sachdev-Ye-Kitaev} model.
\newblock {\em Journal of High Energy Physics}, 2016(4):1--25, 2016.

\bibitem{gross2017generalization}
David~J Gross and Vladimir Rosenhaus.
\newblock A generalization of {Sachdev-Ye-Kitaev}.
\newblock {\em Journal of High Energy Physics}, 2017(2):1--38, 2017.

\bibitem{chowdhury2022sachdevreview}
Debanjan Chowdhury, Antoine Georges, Olivier Parcollet, and Subir Sachdev.
\newblock {Sachdev-Ye-Kitaev} models and beyond: Window into non-fermi liquids.
\newblock {\em Reviews of Modern Physics}, 94(3):035004, 2022.

\bibitem{sarosi2017adsreview}
G{\'a}bor S{\'a}rosi.
\newblock Ads$_2$ holography and the {SYK} model.
\newblock {\em arXiv preprint arXiv:1711.08482}, 2017.

\bibitem{rosenhaus2019introductionreview}
Vladimir Rosenhaus.
\newblock An introduction to the {SYK} model.
\newblock {\em Journal of Physics A: Mathematical and Theoretical}, 52(32):323001, 2019.

\bibitem{huang2019eigenstate}
Yichen Huang and Yingfei Gu.
\newblock Eigenstate entanglement in the {Sachdev-Ye-Kitaev} model.
\newblock {\em Physical Review D}, 100(4):041901, 2019.

\bibitem{passetti2023can}
Giacomo Passetti, Damian Hofmann, Pit Neitemeier, Lukas Grunwald, Michael~A Sentef, and Dante~M Kennes.
\newblock Can neural quantum states learn volume-law ground states?
\newblock {\em Physical Review Letters}, 131(3):036502, 2023.

\bibitem{cerezo2021variational}
Marco Cerezo, Andrew Arrasmith, Ryan Babbush, Simon~C Benjamin, Suguru Endo, Keisuke Fujii, Jarrod~R McClean, Kosuke Mitarai, Xiao Yuan, Lukasz Cincio, et~al.
\newblock Variational quantum algorithms.
\newblock {\em Nature Reviews Physics}, 3(9):625--644, 2021.

\bibitem{mcclean2016theory}
Jarrod~R McClean, Jonathan Romero, Ryan Babbush, and Al{\'a}n Aspuru-Guzik.
\newblock The theory of variational hybrid quantum-classical algorithms.
\newblock {\em New Journal of Physics}, 18(2):023023, 2016.

\bibitem{pikulin2017black}
DI~Pikulin and M~Franz.
\newblock Black hole on a chip: Proposal for a physical realization of the sachdev-ye-kitaev model in a solid-state system.
\newblock {\em Physical Review X}, 7(3):031006, 2017.

\bibitem{cao2020towards}
Ye~Cao, Yi-Neng Zhou, Ting-Ting Shi, and Wei Zhang.
\newblock Towards quantum simulation of sachdev-ye-kitaev model.
\newblock {\em Science Bulletin}, 65(14):1170--1176, 2020.

\bibitem{luo2019quantum}
Zhihuang Luo, Yi-Zhuang You, Jun Li, Chao-Ming Jian, Dawei Lu, Cenke Xu, Bei Zeng, and Raymond Laflamme.
\newblock Quantum simulation of the non-fermi-liquid state of {Sachdev-Ye-Kitaev} model.
\newblock {\em npj Quantum Information}, 5(1):53, 2019.

\bibitem{garcia2017digital}
Laura Garc{\'\i}a-{\'A}lvarez, IL~Egusquiza, Lucas Lamata, Adolfo Del~Campo, Julian Sonner, and Enrique Solano.
\newblock Digital quantum simulation of minimal ads/cft.
\newblock {\em Physical review letters}, 119(4):040501, 2017.

\bibitem{lykken2024long}
Joseph~D Lykken, Daniel Jafferis, Alexander Zlokapa, David~K Kolchmeyer, Samantha~I Davis, Hartmut Neven, and Maria Spiropulu.
\newblock Long-range wormhole teleportation.
\newblock {\em arXiv preprint arXiv:2405.07876}, 2024.

\bibitem{su2020variational}
Vincent~Paul Su.
\newblock Variational preparation of the thermofield double state of the sachdev-ye-kitaev model.
\newblock {\em Physical Review A}, 104(1):012427, 2021.

\bibitem{kim2021entanglement}
Joonho Kim and Yaron Oz.
\newblock Entanglement diagnostics for efficient vqa optimization.
\newblock {\em Journal of Statistical Mechanics: Theory and Experiment}, 2022(7):073101, 2022.

\bibitem{araz2024thermal}
Jack~Y Araz, Raghav~G Jha, Felix Ringer, and Bharath Sambasivam.
\newblock Thermal state preparation of the {SYK} model using a variational quantum algorithm.
\newblock {\em arXiv preprint arXiv:2406.15545}, 2024.

\bibitem{asaduzzaman2023model}
Muhammad Asaduzzaman, Raghav~G Jha, and Bharath Sambasivam.
\newblock Sachdev-ye-kitaev model on a noisy quantum computer.
\newblock {\em Physical Review D}, 109(10):105002, 2024.

\bibitem{goodfellow2016deep}
Ian Goodfellow.
\newblock Deep learning, 2016.

\bibitem{patel2024curriculum}
Yash~J Patel, Akash Kundu, Mateusz Ostaszewski, Xavier Bonet-Monroig, Vedran Dunjko, and Onur Danaci.
\newblock Curriculum reinforcement learning for quantum architecture search under hardware errors.
\newblock In {\em The Twelfth International Conference on Learning Representations}.

\bibitem{kundu2024enhancing}
Akash Kundu, Przemys{\l}aw Bede{\l}ek, Mateusz Ostaszewski, Onur Danaci, Yash~J Patel, Vedran Dunjko, and Jaros{\l}aw~A Miszczak.
\newblock Enhancing variational quantum state diagonalization using reinforcement learning techniques.
\newblock {\em New Journal of Physics}, 26(1):013034, 2024.

\bibitem{patel2024reinforcement}
Yash~J Patel, Sofiene Jerbi, Thomas B{\"a}ck, and Vedran Dunjko.
\newblock Reinforcement learning assisted recursive {QAOA}.
\newblock {\em EPJ Quantum Technology}, 11(1):6, 2024.

\bibitem{ostaszewski2021reinforcement}
Mateusz Ostaszewski, Lea~M Trenkwalder, Wojciech Masarczyk, Eleanor Scerri, and Vedran Dunjko.
\newblock Reinforcement learning for optimization of variational quantum circuit architectures.
\newblock {\em Advances in Neural Information Processing Systems}, 34:18182--18194, 2021.

\bibitem{kuo2021quantum}
En-Jui Kuo, Yao-Lung~L Fang, and Samuel Yen-Chi Chen.
\newblock Quantum architecture search via deep reinforcement learning.
\newblock {\em arXiv preprint arXiv:2104.07715}, 2021.

\bibitem{moro2021quantum}
Lorenzo Moro, Matteo~GA Paris, Marcello Restelli, and Enrico Prati.
\newblock Quantum compiling by deep reinforcement learning.
\newblock {\em Communications Physics}, 4(1):178, 2021.

\bibitem{erdman2024artificially}
Paolo~A Erdman, Robert Czupryniak, Bibek Bhandari, Andrew~N Jordan, Frank Noé, Jens Eisert, and Giacomo Guarnieri.
\newblock Artificially intelligent maxwell’s demon for optimal control of open quantum systems.
\newblock {\em Quantum Science and Technology}, 10(2):025047, mar 2025.

\bibitem{kundu2024reinforcementthesis}
Akash Kundu.
\newblock Reinforcement learning-assisted quantum architecture search for variational quantum algorithms.
\newblock {\em arXiv preprint arXiv:2402.13754}, 2024.

\bibitem{sazli2006brief}
Murat~H Sazl{\i}.
\newblock A brief review of feed-forward neural networks.
\newblock {\em Communications Faculty of Sciences University of Ankara Series A2-A3 Physical Sciences and Engineering}, 50(01), 2006.

\bibitem{svozil1997introduction}
Daniel Svozil, Vladimir Kvasnicka, and Jiri Pospichal.
\newblock Introduction to multi-layer feed-forward neural networks.
\newblock {\em Chemometrics and intelligent laboratory systems}, 39(1):43--62, 1997.

\bibitem{garcia2021sparse}
Antonio~M Garc{\'\i}a-Garc{\'\i}a, Yiyang Jia, Dario Rosa, and Jacobus~JM Verbaarschot.
\newblock Sparse {Sachdev-Ye-Kitaev} model, quantum chaos, and gravity duals.
\newblock {\em Physical Review D}, 103(10):106002, 2021.

\bibitem{xu2020sparse}
Shenglong Xu, Leonard Susskind, Yuan Su, and Brian Swingle.
\newblock A sparse model of quantum holography.
\newblock {\em arXiv preprint arXiv:2008.02303}, 2020.

\bibitem{selisko2023extending}
Johannes Selisko, Maximilian Amsler, Thomas Hammerschmidt, Ralf Drautz, and Thomas Eckl.
\newblock Extending the variational quantum eigensolver to finite temperatures.
\newblock {\em Quantum Science and Technology}, 9(1):015026, 2023.

\bibitem{herasymenko2021diagrammatic}
Y~Herasymenko and TE~O'Brien.
\newblock A diagrammatic approach to variational quantum ansatz construction.
\newblock {\em Quantum}, 5:596, 2021.

\bibitem{kundu2024kanqas}
Akash Kundu, Aritra Sarkar, and Abhishek Sadhu.
\newblock {KANQAS}: {Kolmogorov-Arnold} network for quantum architecture search.
\newblock {\em EPJ Quantum Technology}, 11(1):76, 2024.

\bibitem{fosel2021quantum}
Thomas F{\"o}sel, Murphy~Yuezhen Niu, Florian Marquardt, and Li~Li.
\newblock Quantum circuit optimization with deep reinforcement learning.
\newblock {\em arXiv preprint arXiv:2103.07585}, 2021.

\bibitem{jin2014flattened}
Jonghoon Jin, Aysegul Dundar, and Eugenio Culurciello.
\newblock Flattened convolutional neural networks for feedforward acceleration.
\newblock {\em arXiv preprint arXiv:1412.5474}, 2014.

\bibitem{rao2019natural}
Delip Rao and Brian McMahan.
\newblock {\em Natural language processing with PyTorch: build intelligent language applications using deep learning}.
\newblock " O'Reilly Media, Inc.", 2019.

\bibitem{suzuki2021qulacs}
Yasunari Suzuki, Yoshiaki Kawase, Yuya Masumura, Yuria Hiraga, Masahiro Nakadai, Jiabao Chen, Ken~M Nakanishi, Kosuke Mitarai, Ryosuke Imai, Shiro Tamiya, et~al.
\newblock Qulacs: a fast and versatile quantum circuit simulator for research purpose.
\newblock {\em Quantum}, 5:559, 2021.

\bibitem{powell2007view}
Michael~JD Powell.
\newblock A view of algorithms for optimization without derivatives.
\newblock {\em Mathematics Today-Bulletin of the Institute of Mathematics and its Applications}, 43(5):170--174, 2007.

\bibitem{kingma2014adam}
Diederik~P Kingma.
\newblock {ADAM}: A method for stochastic optimization.
\newblock {\em arXiv preprint arXiv:1412.6980}, 2014.

\bibitem{guo2022quantum}
Yuchen Guo and Shuo Yang.
\newblock Quantum error mitigation via matrix product operators.
\newblock {\em PRX Quantum}, 3(4):040313, 2022.

\bibitem{filippov2023scalable}
Sergei Filippov, Matea Leahy, Matteo~AC Rossi, and Guillermo Garc{\'\i}a-P{\'e}rez.
\newblock Scalable tensor-network error mitigation for near-term quantum computing.
\newblock {\em arXiv preprint arXiv:2307.11740}, 2023.

\bibitem{botelho2022error}
Ludmila Botelho, Adam Glos, Akash Kundu, Jaros{\l}aw~Adam Miszczak, {\"O}zlem Salehi, and Zolt{\'a}n Zimbor{\'a}s.
\newblock Error mitigation for variational quantum algorithms through mid-circuit measurements.
\newblock {\em Physical Review A}, 105(2):022441, 2022.

\bibitem{robbiati2023real}
Matteo Robbiati, Alejandro Sopena, Andrea Papaluca, and Stefano Carrazza.
\newblock Real-time error mitigation for variational optimization on quantum hardware.
\newblock {\em arXiv preprint arXiv:2311.05680}, 2023.

\bibitem{cai2023quantum}
Zhenyu Cai, Ryan Babbush, Simon~C Benjamin, Suguru Endo, William~J Huggins, Ying Li, Jarrod~R McClean, and Thomas~E O’Brien.
\newblock Quantum error mitigation.
\newblock {\em Reviews of Modern Physics}, 95(4):045005, 2023.

\bibitem{kundu2024easy}
Akash Kundu and Leopoldo Sarra.
\newblock Reinforcement learning with learned gadgets to tackle hard quantum problems on real hardware.
\newblock {\em arXiv preprint 2411.00230}, 2025.

\bibitem{tan2018survey}
Chuanqi Tan, Fuchun Sun, Tao Kong, Wenchang Zhang, Chao Yang, and Chunfang Liu.
\newblock A survey on deep transfer learning.
\newblock In {\em Artificial Neural Networks and Machine Learning--ICANN 2018: 27th International Conference on Artificial Neural Networks, Rhodes, Greece, October 4-7, 2018, Proceedings, Part III 27}, pages 270--279. Springer, 2018.

\bibitem{sadhu2024quantum}
Abhishek Sadhu, Aritra Sarkar, and Akash Kundu.
\newblock A quantum information theoretic analysis of reinforcement learning-assisted quantum architecture search.
\newblock {\em Quantum Machine Intelligence}, 6(2):49, 2024.

\bibitem{cai2021multi}
Zhenyu Cai.
\newblock Multi-exponential error extrapolation and combining error mitigation techniques for nisq applications.
\newblock {\em npj Quantum Information}, 7(1):80, 2021.

\bibitem{endo2018practical}
Suguru Endo, Simon~C Benjamin, and Ying Li.
\newblock Practical quantum error mitigation for near-future applications.
\newblock {\em Physical Review X}, 8(3):031027, 2018.

\bibitem{kelley1928crossroads}
Truman~Lee Kelley.
\newblock {\em Crossroads in the mind of man: A study of differentiable mental abilities}.
\newblock Stanford university press, 1928.

\bibitem{hurst1995characteristic}
Simon Hurst.
\newblock {\em The characteristic function of the Student t distribution}.
\newblock Centre for Mathematics and its Applications, School of Mathematical Sciences~…, 1995.

\bibitem{larose2019variational}
Ryan LaRose, Arkin Tikku, {\'E}tude O’Neel-Judy, Lukasz Cincio, and Patrick~J Coles.
\newblock Variational quantum state diagonalization.
\newblock {\em npj Quantum Information}, 5(1):57, 2019.

\bibitem{huang2025direct}
Yulei Huang, Liangyu Che, Chao Wei, Feng Xu, Xinfang Nie, Jun Li, Dawei Lu, and Tao Xin.
\newblock Direct entanglement detection of quantum systems using machine learning.
\newblock {\em npj Quantum Information}, 11(1):29, 2025.

\bibitem{wang2023quantum}
Youle Wang, Benchi Zhao, and Xin Wang.
\newblock Quantum algorithms for estimating quantum entropies.
\newblock {\em Physical Review Applied}, 19(4):044041, 2023.

\bibitem{czarnik2021error}
Piotr Czarnik, Andrew Arrasmith, Patrick~J Coles, and Lukasz Cincio.
\newblock Error mitigation with clifford quantum-circuit data.
\newblock {\em Quantum}, 5:592, 2021.

\bibitem{verdon2019quantum}
Guillaume Verdon, Jacob Marks, Sasha Nanda, Stefan Leichenauer, and Jack Hidary.
\newblock Quantum hamiltonian-based models and the variational quantum thermalizer algorithm.
\newblock {\em arXiv preprint arXiv:1910.02071}, 2019.

\bibitem{sutton2018reinforcement}
Richard~S Sutton.
\newblock Reinforcement learning: An introduction.
\newblock {\em A Bradford Book}, 2018.

\bibitem{van2016deep}
Hado Van~Hasselt, Arthur Guez, and David Silver.
\newblock Deep reinforcement learning with double q-learning.
\newblock In {\em Proceedings of the AAAI conference on artificial intelligence}, volume~30, 2016.

\bibitem{mnih2015human}
Volodymyr Mnih, Koray Kavukcuoglu, David Silver, Andrei~A Rusu, Joel Veness, Marc~G Bellemare, Alex Graves, Martin Riedmiller, Andreas~K Fidjeland, Georg Ostrovski, et~al.
\newblock Human-level control through deep reinforcement learning.
\newblock {\em nature}, 518(7540):529--533, 2015.

\bibitem{mittal2021survey}
Sparsh Mittal et~al.
\newblock A survey of accelerator architectures for 3d convolution neural networks.
\newblock {\em Journal of Systems Architecture}, 115:102041, 2021.

\bibitem{hossain2019classification}
Md~Anwar Hossain and Md~Shahriar~Alam Sajib.
\newblock Classification of image using convolutional neural network ({CNN}).
\newblock {\em Global Journal of Computer Science and Technology}, 19(2):13--14, 2019.

\bibitem{zheng2016good}
Liang Zheng, Yali Zhao, Shengjin Wang, Jingdong Wang, and Qi~Tian.
\newblock Good practice in {CNN} feature transfer.
\newblock {\em arXiv preprint arXiv:1604.00133}, 2016.

\bibitem{zhu2022lite}
Maochang Zhu, Sheng Bin, and Gengxin Sun.
\newblock {Lite-3DCNN} combined with attention mechanism for complex human movement recognition.
\newblock {\em Computational Intelligence and Neuroscience}, 2022(1):4816549, 2022.

\bibitem{siddiqui2021progressive}
Zahid~Ali Siddiqui and Unsang Park.
\newblock Progressive convolutional neural network for incremental learning.
\newblock {\em Electronics}, 10(16):1879, 2021.

\end{thebibliography}

\appendix

{\section{In depth analysis of CNOT scaling}\label{appendix:scaling_of_snot_analysis}
In Fig.~\ref{fig:cnot_scaling}, we demonstrate that the number of CNOT gates scales polynomially with qubit count, approximately following the cubic trend
\begin{equation}
    y = 2.06N_q^3 - 27.73N_q^2 + 123.54N_q - 168.71N_q,
\end{equation}
where $N_q$ is the number of qubits. This represents a substantial advantage over the exponential scaling of real-time dynamics
\begin{equation}
    y = 14.839e^{0.890N_q} - 379.902.
\end{equation}
\begin{figure}[h!]
    \centering
    \includegraphics[width=\linewidth]{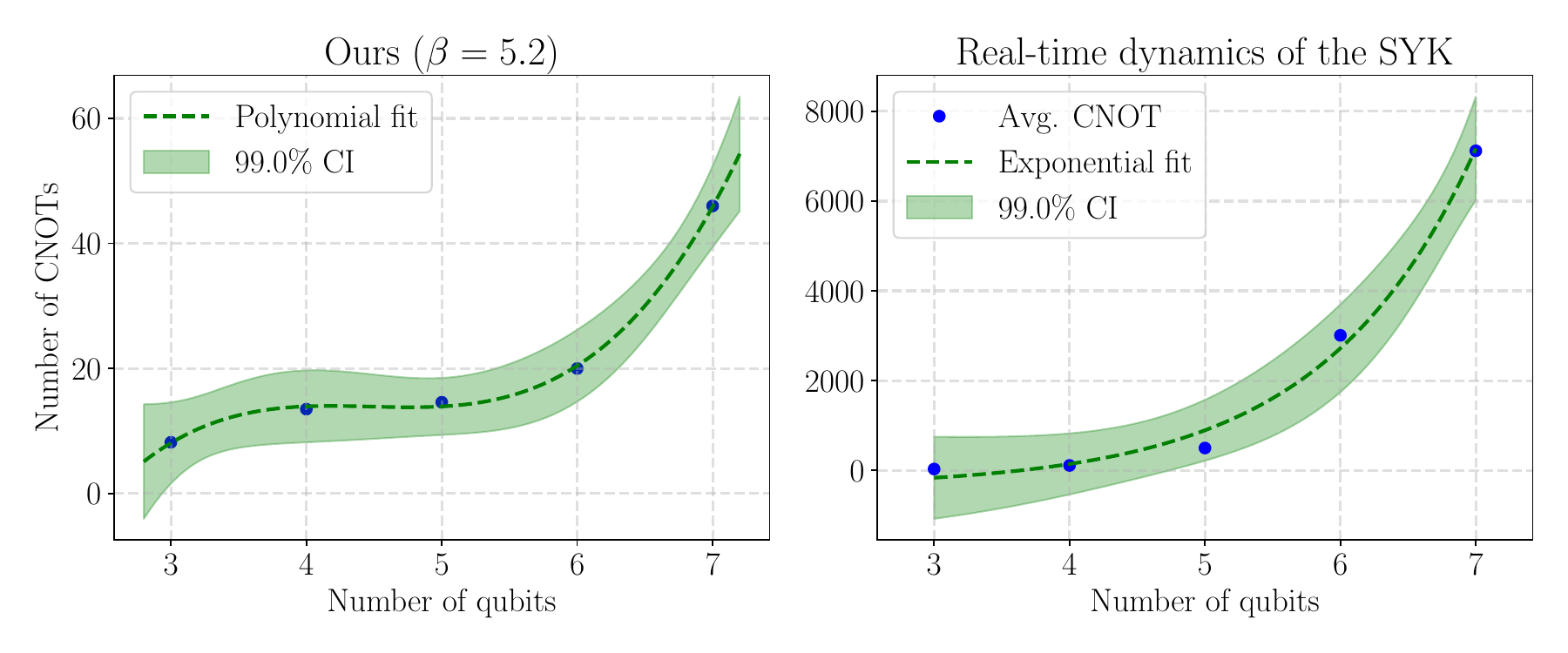}
    \caption{{\textbf{Scaling of CNOT gate counts with qubit number for our method (polynomial fit) versus real-time dynamics (exponential fit)}. The cubic scaling of our approach ($y = 2.06N_q^3 - 27.73N_q^2 + 123.54N_q - 168.71$) offers a substantial advantage over the exponential growth observed in real-time dynamics ($y = 14.839e^{0.890N_q} - 379.902$), particularly for larger systems. Fitted curves and data points are shown for 3–10 qubits, highlighting the divergence in gate requirements as system size increases. Analytical Trotterization results for 10 qubits are also indicated for comparison.}}
    \label{fig:cnot_scaling}
\end{figure}
This scaling difference becomes crucial for larger systems: our method maintains $\mathcal{O}(N_q^3)$ CNOT growth with qubit number, while real-time dynamics exhibits $\mathcal{O}(e^{0.890N_q})$ scaling. The fitted curves reveal comparable CNOT counts for small systems: $8$ (ours) vs.\ $30$ (real-time) for 3-qubit systems and $13$ vs.\ $110$ for 4-qubit systems. However, the counts diverge dramatically as $N$ increases. This polynomial growth pattern suggests feasibility for larger SYK simulations on near-term hardware. Notably, the exponential scaling equation overestimates gate counts at higher qubit numbers: for 10-qubit systems, it predicts $\sim\!230000$ gates, while analytical Trotterization yields $19984$ CNOTs per layer~\cite{asaduzzaman2023model}. With two layers ($39968$ total), this remains significantly below the exponential prediction. This discrepancy arises from limited data points during curve fitting.
}

{The confidence intervals (CI) visualized in the green shaded regions were computed using the delta method~\cite{kelley1928crossroads} for uncertainty propagation in nonlinear models. For the exponential model $f(x) = a \cdot e^{bx} + c$, the Jacobian matrix $J$ was constructed containing partial derivatives to each parameter ($\partial f/\partial a = e^{bx}$, $\partial f/\partial b = ax \cdot e^{bx}$, and $\partial f/\partial c = 1$). The variance of the prediction at each point $x_i$ was then calculated as $\sigma_{y_i}^2 = J_i^T \Sigma J_i$, where $\Sigma$ is the parameter covariance matrix obtained from the curve-fitting procedure. For a confidence level $\alpha$ (0.05 for 95\% confidence), the intervals were computed as $\hat{y}i \pm t{n-p,1-\alpha/2} \cdot \sqrt{\sigma_{y_i}^2}$, where $t_{n-p,1-\alpha/2}$ is the critical value from Student's t-distribution~\cite{hurst1995characteristic} with $n-p$ degrees of freedom ($n$ being the number of data points and $p$ the number of parameters). The widening confidence bands at higher qubit counts reflect the increasing uncertainty in predictions as we extrapolate beyond the measured data range, with the exponential model showing particularly large uncertainties due to the compounding effect of parameter uncertainty in the exponent.}

{\section{Resource requirements and scalability of entropy measurement from PQC$_1$}\label{appndix:entropy_measurement}
Accurate entropy measurement in quantum systems necessitates a careful balance between precision and resource efficiency. Traditional approaches such as {quantum state tomography} reconstruct the full density matrix through repeated state preparation and measurement but suffer from exponential scaling, requiring $O(3^n)$ measurements for $n$ qubits. This limits practical applications to systems with $n \gtrsim 10$ qubits. Modern methodologies achieve polynomial efficiency under specific constraints:}

{Variational quantum diagonalization employs parameterized quantum circuits optimized to estimate von Neumann entropy with $O(\mathrm{poly}(n))$ measurements~\cite{larose2019variational,kundu2024enhancing}. While measurement-efficient, these methods incur classical optimization overhead that scales with circuit depth. {Classical shadows} leverage randomized single-qubit measurements to estimate the second R\'enyi entropy with $O(\log n)$ measurements per state copy~\cite{huang2025direct}, though they provide limited insight into mixed-state von Neumann entropy. Hybrid approaches combining {machine learning methods} with local measurements enable entanglement entropy prediction for systems up to $\sim 100$ qubits~\cite{huang2025direct}, albeit with assumptions about spatial entanglement structure.}

{Scalability depends critically on state purity. For low-rank states (e.g., pure or low-temperature thermal states), variational and shadow-based methods exhibit polynomial scaling, demonstrated experimentally for $n \leq 50$ qubits~\cite{wang2023quantum}. In contrast, high-entropy full-rank systems face fundamental $O(2^n)$ measurement bottlenecks due to Hilbert space growth. All techniques require $O(1/\epsilon^2)$ measurements for precision $\epsilon$, with error mitigation strategies~\cite{botelho2022error,czarnik2021error} becoming essential on noisy hardware. Recent advances include {reinforcement learning-optimized protocols} that adaptively minimize measurement rounds~\cite{kundu2024enhancing} and {symmetry-aware shadow estimation} reducing sampling costs by 60\% for states with conserved quantities~\cite{wang2023quantum}. While polynomial-scaling methods enable practical entropy estimation for intermediate-scale quantum devices ($n \sim 100$), challenges persist for highly mixed states and systems with non-Markovian noise. Hybrid quantum-classical algorithms leveraging problem-specific structure currently offer the most viable path forward, though fundamental limitations remain for generic high-entropy systems.}

{\section{Elaboration on variational thermal state preparation}\label{appendix:why_vqtsp_works}
The task of variational thermalization is: given a Hamiltonian $H$ and a target temperature $T=1/\beta$, we need to prepare $\rho_\beta$ (the thermal state), which is expressed by
\begin{equation}
    \rho_\beta = \frac{e^{-\beta H}}{\operatorname{tr}(e^{-\beta H})},
    \label{eq:thermal_state}
\end{equation}
where the partition function $Z_\beta = \operatorname{tr}(e^{-\beta H})$. To implement this in a variational quantum algorithm, we consider a quantum ansatz state $\rho_{\theta\phi}$ and define the relative entropy between the ansatz state and the thermal state as
\begin{equation}
    D(\rho_{\theta\phi}||\rho_\beta) = -S(\rho_{\theta\phi}) - \operatorname{tr}(\rho_{\theta\phi}\log\rho_\beta),
\end{equation}
which can be rewritten using Eq.~\ref{eq:thermal_state} as
\begin{equation}
    D(\rho_{\theta\phi}||\rho_\beta) = -S(\rho_{\theta\phi}) + \beta \operatorname{tr}(\rho_{\theta\phi}H) + \log Z_\beta.
\end{equation}
This is known as the relative quantum entropy. Using the free energy definition $F(\cdot) \equiv \operatorname{tr}((\cdot)H) - \frac{S(\cdot)}{\beta}$, we can rewrite this as
\begin{equation}
    D(\rho_{\theta\phi}||\rho_\beta) = \beta F(\rho_{\theta\phi}) - \beta F(\rho_\beta).
    \label{eq:free_energy_relation}
\end{equation}
Thus, $\rho_{\theta\phi}$ represents the thermal state $\rho_\beta$ iff $F(\rho_{\theta\phi}) = F(\rho_\beta)$ and $\rho_{\theta\phi} = \rho_\beta$, i.e.,
\begin{equation}
    D(\rho_{\theta\phi}||\rho_\beta) = 0.
    \label{eq:minimization1}
\end{equation}
Therefore, for a variational thermal state ansatz, minimizing the free energy
\begin{equation}
    C(\rho_{\theta\phi}) = \beta F(\rho_{\theta\phi}) = \beta \operatorname{tr}(\rho_{\theta\phi} H) - S(\rho_{\theta\phi})
    \label{eq:minimization2}
\end{equation}
yields optimal parameters where $\rho_{\theta\phi} = \rho_\beta$.}

{In a state-of-the-art~\cite{verdon2019quantum} implementation, entropy calculation occurs classically while energy expectation is on a quantum computer. To compute both on a quantum computer~\cite{selisko2023extending}, as we do in our case:
\begin{itemize}
    \item $\rho_\theta$ is generated via a parameterized quantum circuit $\mathrm{PQC}_1(\theta)$. We use $\rho_\theta$ to calculate the entropy $S(\rho_{\theta\phi})$. 
    \item $\rho_{\theta\phi}$ is obtained by initialized another parameterized quantum circuit $\mathrm{PQC}_2(\phi)$ using $\rho_\theta$. Then $\rho_{\theta\phi}$ is used to calculate the Hamiltonian expectation value $E(\rho_{\theta\phi})$.
    \item $E(\rho_{\theta\phi})$ and $S(\rho_{\theta\phi})$ compute the free energy $\beta F(\rho_{\theta\phi})$ for specific $\beta$.
\end{itemize}
Equations~\ref{eq:minimization1} and~\ref{eq:minimization2} show that $ D(\rho_{\theta\phi}||\rho_\beta) = 0$ implies $\rho_{\theta\phi} = \rho_\theta = \rho_\beta$. Hence, during optimization $\rho_\theta \neq \rho_{\theta\phi}$ but when the loss $C(\rho_{\theta\phi})$ is minimized we get $\rho_\theta = \rho_{\theta\phi} = \rho_\beta$.}

\section{Reinforcement learning framework}\label{appndix:rl_framework}

In the context of Reinforcement Learning (RL), an agent interacts with its environment to learn an optimal policy through a process of trial and error~\cite{sutton2018reinforcement}. This interaction can be formally modeled as a Markov Decision Process (MDP), defined by the tuple $(S,A,P,R)$, where:

\begin{itemize}
    \item $S$ represents the state space
    \item $A$ represents the action space
    \item $P : S \times S \times A \rightarrow [0, 1]$ defines the transition dynamics
    \item $R : S \times A \rightarrow \mathbb{R}$ describes the reward function of the environment
\end{itemize}

In our work, we consider both the action space $A$ and the state space $S$ to be finite and discrete sets.

\paragraph{Agent behavior and performance metrics}

The agent's behavior within the environment is governed by a stochastic policy $\pi(a|s) : S \times A \rightarrow [0, 1]$, where $a \in A$ and $s \in S$. To assess the agent's performance, we use a metric called the \emph{return}, which is calculated as a discounted sum:

\begin{equation}
    G(\tau)=\sum_{j=0}^{T-1} r_j \gamma^{j+1}
\end{equation}

Here, $\tau=\left(s_0, a_0, r_0, \ldots, s_{T-1}, a_{T-1}, r_{T-1}\right) \in(S \times A \times \mathbb{R})^T$ represents the interaction sequence $T$ is a fixed length called the horizon, and $\gamma$ is an environment-specific discount factor. The primary objective of the agent is to determine the optimal policy that maximizes the expected return.

\paragraph{Function approximation and algorithm choice} To handle large, unknown environments where the agent must adapt to various situations and develop multiple strategies simultaneously, we employ highly expressive function approximations. Specifically, we use deep neural networks to parametrize the agent's policy $\pi$.

For our implementation, we have chosen to use a Double Deep Q-Network (DDQN)~\cite{van2016deep}, which is an extension of the standard Deep Q-Network (DQN)~\cite{mnih2015human}. The DDQN algorithm utilizes two neural networks to enhance the stability of Q-value predictions for each state-action pair.

\paragraph{State and action space representation} In our specific application:
\begin{itemize}
    \item The state space is represented as an ordered list of layers, each composed of a single depth of the quantum circuit.
    \item The action space is defined by a list of four numbers, corresponding to the quantum gates \texttt{RX}, \texttt{RY}, \texttt{RZ}, and CNOT.
\end{itemize}


\section{Tensor-based encoding of quantum circuits}\label{appndix:tensor_encoding}

We adopt a binary encoding scheme, as introduced in~\cite{patel2024curriculum,kundu2024enhancing}, to capture the structure of PQC$_2$, focusing on the order and arrangement of gates to provide the agent with a complete circuit description. Unlike previous approaches, which flattened the 3D encoding into one dimension for use in feedforward neural networks, our method leverages the full spatial structure of the encoding by employing a 3D convolutional neural network (CNN). This allows us to extract richer spatial features from the encoded representation as shown in the Appendix~\ref{appndix:CNN_vs_FNN}.
. 

\begin{figure}[h!]
    \centering
    \includegraphics[width=\linewidth]{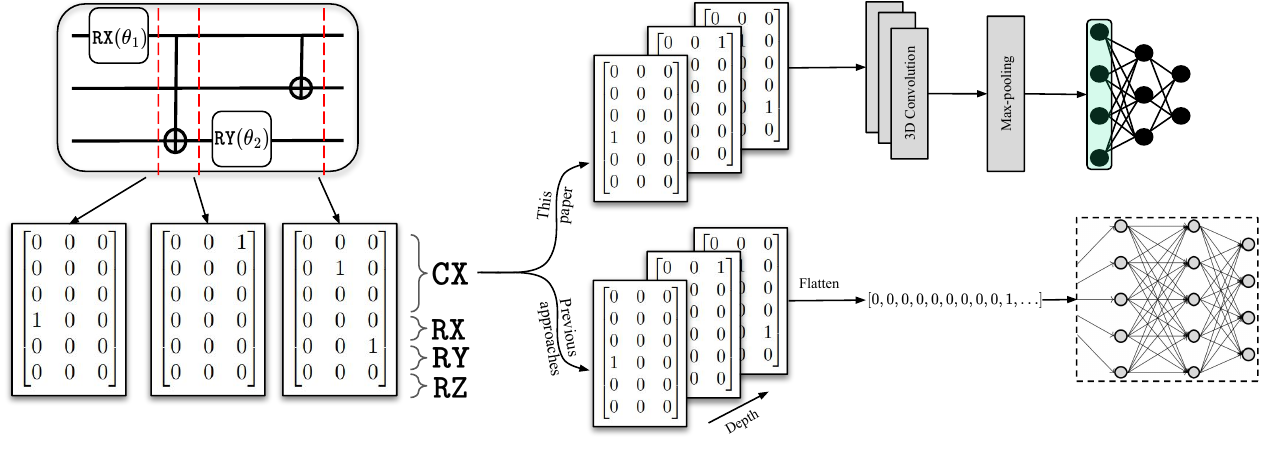}
    \caption{\small\textbf{The 3D tensor encoding of the PQC$_2$ circuit structure}. The representation captures gate number, order, and arrangement, enabling the use of 3D-CNN for richer spatial feature extraction, in contrast to flattened 1D encodings used with feedforward networks in previous approaches.}
    \label{fig:encoding}
\end{figure}

To construct the tensor encoding, we first define the hyperparameter $D_{\text{max}}$, which sets the maximum number of allowed gates (or actions) in an episode. A \textit{moment} in a PQC corresponds to all gates that can be executed simultaneously, thereby defining the circuit's depth. 

The PQCs are represented as 3D tensors initialized at each episode with an empty circuit of depth $D_{\text{max}}$. This tensor has dimensions $\left[D_{\text{max}} \times \left((N+N_\text{1q})\times N\right)\right]$, where $N$ represents the number of qubits, and $N_\text{1q}$ denotes the number of single-qubit gates. Each slice of the tensor has $N$ rows for the control and target qubit positions of CNOT gates, followed by rows indicating the positions of single-qubit gates. For instance, as shown in Fig.~\ref{fig:encoding}, there are 3 rows for three spatial rotations, \texttt{RX}, \texttt{RY} and \texttt{RZ}.

\section{CNN outperforms FNN}\label{appndix:CNN_vs_FNN}

We utilize 3D convolution, whose theory can be found elaborately in~\cite{mittal2021survey, goodfellow2016deep}. One layer of 3D-CNN is defined as
\begin{lstlisting}[language=Python, caption=\textbf{One layer of 3D-CNN}, label = listing:3D-CNN_structure]
for channel in channels_list:
    layer_list.append(Conv3d(input_channel, channel, kernel_size=(3,3,3), stride=(1,1,1), padding=(1,1,1)))
    layer_list.append(LeakyReLU())
    layer_list.append(MaxPool3d(kernel_size=(1,1,1), stride=(2,2,2), padding=0, dilation=1))
    input_channel = channel
\end{lstlisting}
In tackling the $N=8$ SYK Hamiltonian, we consider \texttt{channels\_list} in \ref{listing:3D-CNN_structure} as follows.
For (1) \textit{4 layers}: [32, 64, 128, 256], for (2) \textit{5 layers}: [32, 64, 128, 256, 512], for (3) \textit{6 layers}: [32, 64, 128, 256, 512, 1024] and for (4) \textit{7 layers}: [32, 64, 128, 256, 512, 1024, 2048]. 

Following the standard approach in convolutional neural networks~\cite{hossain2019classification,zheng2016good,zhu2022lite,siddiqui2021progressive}, the architecture presented progressively increases the number of channels. Starting from 32 channels and doubling them at each layer. This is a common strategy to capture more intricate patterns in the data, which allows the network to learn increasingly complex features at deeper layers. The initial layers capture low-level features, while deeper layers capture high-level features. In the language of quantum architecture search the input to the 3DCNN is comprised of a 3D tensor whose axes represent quantum circuit depth, type of gates in the circuit, and the number of qubits; the object of the quantum circuit is to find the ground state of
\begin{figure}[h!]
    \centering
    \includegraphics[width=\linewidth]{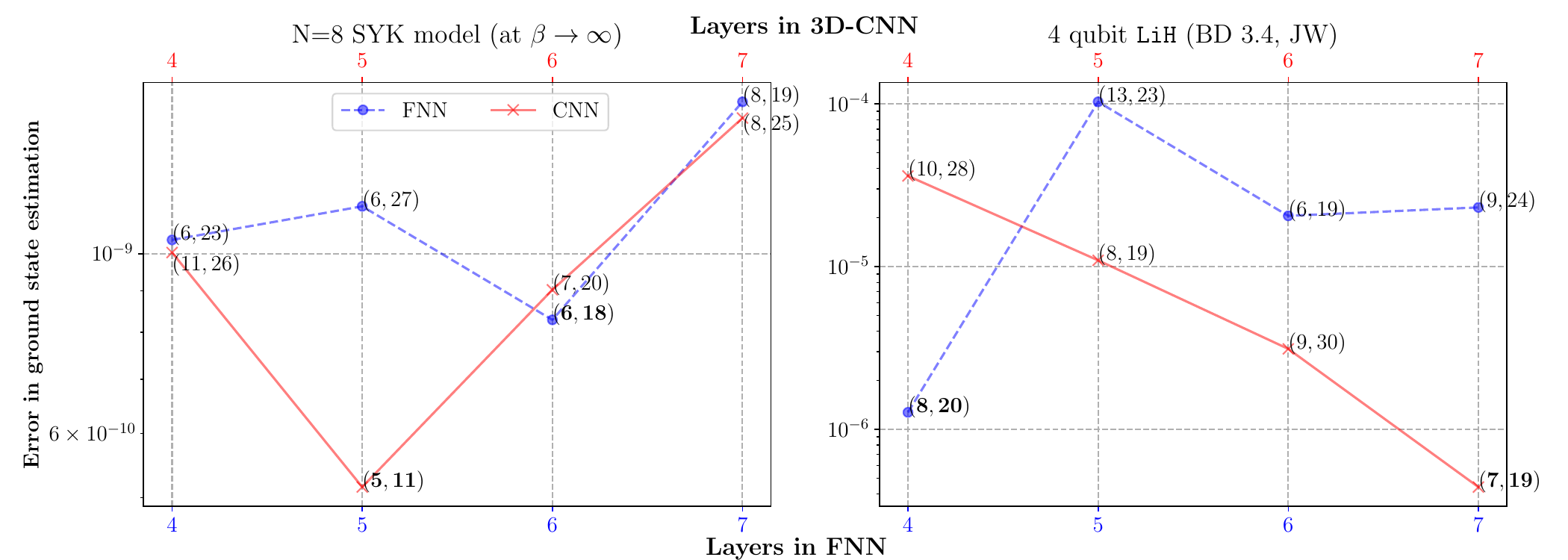}
    \caption{\small\textbf{Convolutional neural network (CNN) achieves a lower error in finding the ground state of the SYK model at $N=8$ and finding the ground state of \texttt{LiH} molecule compared to feedforward neural network (FNN)}. Along with $N=8$ SYK Hamiltonian, we consider a well-known \texttt{LiH} chemical Hamiltonian to show the advantages of utilizing CNN over FNN in the RL framework. In the plot, by ($a$,$b$) we represent a CNOT and $b$ 1-qubit rotations (i.e., total number of \texttt{RX}, \texttt{RY} and \texttt{RZ}).
    The CNN not only achieved much lower error in comparison to the FNN but did it with a fewer number of 1- and 2-qubit gates for both the problems. }
    \label{fig:cnn_vs_fnn}
\end{figure}
Hamiltonian. Hence the low-level features are defined by the depth, number of gates, and the number of qubits in the circuit, whereas the high-level feature is represented by the combinations of gates in the circuit, the performance metrics such as the objective function, and the reward received from the environment. Hence, increasing the layer of the 3D-CNN not only helps optimize the depth and the number of gates of the circuit using the first few layers, but the network makes sure that these circuits successfully reach the ground truth.
\vspace{2pt}

Whereas one-layer of 1D FNN is defined as:
\begin{lstlisting}[language=Python, caption=\textbf{One layer of FNN}, label = listing:FNN_structure]
for input_n, output_n in zip(neuron_list[:-1], neuron_list[1:]):
    layer_list.append(nn.Linear(input_n, output_n))
    layer_list.append(nn.LeakyReLU())
    layer_list.append(nn.Dropout(p=p))
\end{lstlisting}

In tackling the $N=8$ SYK Hamiltonian, we consider \texttt{channels\_list} in \ref{listing:FNN_structure} as follows.
For (1) \textit{4 layers}: [1000, 1000, 1000, 1000], for (2) \textit{5 layers}: [1000, 1000, 1000, 1000, 1000], for (3) \textit{6 layers}: [1000, 1000, 1000, 1000, 1000, 1000] and for (4) \textit{7 layers}: [1000, 1000, 1000, 1000, 1000, 1000, 1000].
\begin{table}[h!]
\centering
\begin{tabular}{l|cccc}
\hline
Method & Error & Parameters & Depth & CNOT \\
\hline
CRLQAS~\cite{patel2024curriculum}& \(2.6 \times 10^{-6}\) & 29 & 22 & 11 \\
KANQAS~\cite{kundu2024kanqas} & \(1.1 \times 10^{-4}\) & $\mathbf{15}$ & 18 & 11 \\
Ours & $\mathbf{4.42 \times 10^{-7}}$ & 19 & $\mathbf{16}$ & $\mathbf{7}$ \\
\hline
\end{tabular}
\caption{{Comparison of quantum circuit metrics.}}
\label{tab:cnn_vs_fnn_vs_kan}
\end{table}

{Fig.~\ref{fig:cnn_vs_fnn} not only demonstrates the clear advantage of using CNNs over FNNs within the RL framework but also highlights their advantage over state-of-the-art quantum architecture search algorithms, such as KANQAS~\cite{kundu2024kanqas} and CRLQAS~\cite{patel2024curriculum}. Specifically, the proposed approach achieves higher accuracy in finding the ground state of chemical Hamiltonians while utilizing smaller quantum circuits. A detailed comparison of these results is provided in Tab.~\ref{tab:cnn_vs_fnn_vs_kan}.}

\section{Training time}
In Tab.~\ref{tab:resource_time}
we discuss the noiseless/noisy simulation time by various SYK Hamiltonians. All the simulation is done under a computational budget of 48 hours within which the RL-framework completed approximately $5\times10^{3}$ episodes.
\begin{table}[h!]
\centering
\begin{tabular}{c|cccc}
{Majorana fermions ($N$)} & 8 & 10 & 12 & 14 \\
\hline
\hline
{Time (in hours)} & 3.9162 & 24.2600 & 47.4669 & 47.8951 \\
\end{tabular}
\caption{\textbf{The RL-agent training time} to obtain the results presented in Sec.~\ref{sec:results} for various number of Majorana fermions.}
\label{tab:resource_time}
\end{table}

\section{The \texttt{IBM Eagle r3} processor}\label{appndix:QPU}
\begin{figure}[h!]
    \centering    \includegraphics[width=0.8\linewidth]{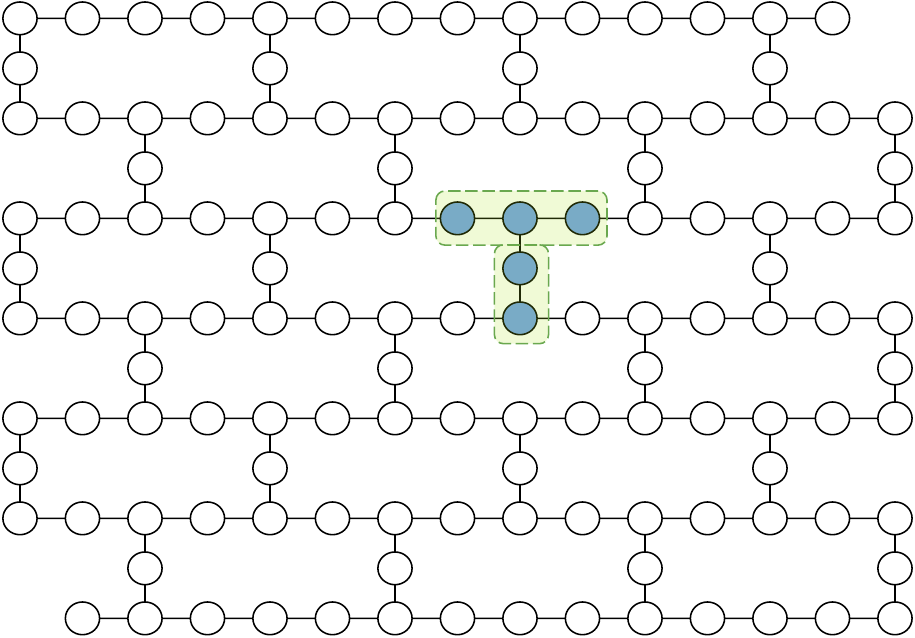}
    \caption{\small\textbf{Topology of the 127-qubit \texttt{IBM Eagle r3} processors used in noisy experiments}. As we run the SYK Hamiltonian for $N=8$ Majorana fermions, in the highlighted part of the topology we show that in this experiment we utilized a T-configuration that minimizes additional SWAP gates required due to the restricted connectivity of physical qubits as noted in ref.~\cite{asaduzzaman2023model}.}
    \label{fig:connectivity}
\end{figure}

The free energy and the entropy is calculated on \texttt{ibm\_kyiv}, \texttt{ibm\_sherbrooke} and \texttt{ibm\_brisbane}. They are all 127 qubit devices with the \texttt{Eagle r3} processor. The qubit connectivity of these devices is shown in Fig.~\ref{fig:connectivity}. All these machines use the same set of basis gates, which consist of
ECR and single-qubit gates \texttt{ID}, \texttt{RZ}, \texttt{SX}, \texttt{X}.

\section{The agent and environment hyperparameters}\label{appndix:agent_hyperparameters}

The hyperparameter tuning of the reinforcement learning agent corresponds to the tuning of the following parameters:
\begin{table}[ht]
\centering
\small
\begin{tabular}{*{21}{c}}
\hline
{Batch size} & {Memory size} & {Dropout} & {Update target net} & {Final $\gamma$} & {$\epsilon$ decay} & {$\epsilon$ min} & {$\epsilon$ restart} \\ \hline
$10^{3}$ & $2\times10^{4}$ & 0 & 500 & $5\times10^{-3}$ & 0.99995 & $5\times10^{-2}$ & 1 \\ \hline
\end{tabular}
\caption{\small The hyperparameters of the reinforcement learning agent.}
\label{table:agent_hyperparameters}
\end{table}
\begin{enumerate}
    \item {Batch size:} The number of experiences sampled from the memory for each training step.
    \item {Memory size:} The total capacity of the experience replay buffer storing past interactions.
    \item {Dropout:} The probability of randomly disabling neurons during training to prevent overfitting.
    \item {Update target net:} The frequency (in steps) at which the target network is updated with the online network's weights.
    \item {Final $\gamma$:} The discount factor determines the importance of future rewards compared to immediate rewards.
    \item {$\epsilon$:} The rate at which the exploration probability decreases in the $\epsilon$-greedy policy.
    \item {$\epsilon$ min:} The minimum value of $\epsilon$, ensuring a baseline level of exploration.
    \item {$\epsilon$ restart:} The initial value of $\epsilon$ at the start of training or after a reset, representing the maximum exploration level.
\end{enumerate}

{\section{Reward function analysis}\label{appndix:rwd_function_analysis}
In this section, we discuss the effect of weight corresponding to free energy ($a$ and the weight corresponding to Hamiltonian expectation ($b$) on the reward function
\begin{equation}
r = 
\begin{cases} 
5, & \text{if } F_t(\beta=\text{const.},\vec{\theta},\vec{\phi}) {\leq} \zeta_F\; \text{and,}\;\textrm{Fid}(\rho({\vec{\phi}})) \geq \zeta_\textrm{Fid.} \\
-5, & \text{if } \textrm{Fid}(\rho({\vec{\phi}})) < \zeta_\textrm{Fid} \;\text{and,}\; \text{step no.} = D_\text{max}, \\
a \times E_\text{term} + b \times \textrm{Fid}_\text{term}, & \text{otherwise},
\end{cases}
\label{case:error_and_fidelity_rwd}
\end{equation}
The analysis in Tab.~\ref{tab:reward_analysis} reveals a nuanced relationship between the precision of free energy estimation and the focus on expectation values. As shown in Table~\ref{tab:reward_analysis}, increasing the weight $a$ generally leads to improved precision in free energy estimation, with the lowest error of $2.15\times10^{-7}$ observed at $a=0.8$ and $b=0.2$.}
\begin{table}[h!]
\centering
\sisetup{
    exponent-product = \ensuremath{\times},
    tight-spacing = true
}
\begin{tabular}{cccc}
\toprule
{a} & {b} & {Free energy error} & {Hamiltonian expectation error} \\
\midrule
0.0 & 1.0 & $5.14\times10^{-6}$ & $7.84\times10^{-4}$ \\
0.2 & 0.8 & $1.64\times10^{-6}$ & $2.67\times10^{-3}$ \\
0.4 & 0.6 & $2.17\times10^{-6}$ & $8.35\times10^{-4}$ \\
0.6 & 0.4 & $1.81\times10^{-6}$ & $2.73\times10^{-3}$ \\
0.8 & 0.2 & $2.15\times10^{-7}$ & $2.53\times10^{-3}$ \\
1.0 & 0.0 & $5.14\times10^{-6}$ & $7.84\times10^{-4}$ \\
\bottomrule
\end{tabular}
\caption{{Free energy and expectation value errors for different reward function weight configurations. Where $a$ = weight corresponding to free energy, $b$ = weight corresponding to Hamiltonian expectation value.}}\label{tab:reward_analysis}
\end{table}

{However, this trend does not continue indefinitely. When $a$ is set to 1.0, effectively eliminating the focus on the Hamiltonian expectation value ($b=0.0$), we observe a decrease in precision, with the free energy error increasing to $5.14\times10^{-6}$. This suggests that maintaining some weight on the expectation value is crucial for optimal performance.}

{These findings highlight the importance of balancing the emphasis between free energy and expectation value in the reward function. While prioritizing free energy estimation (higher $a$ values) generally improves precision, completely neglecting the expectation value component can lead to suboptimal results. Future work could explore finer-grained weight configurations to determine the optimal balance for specific applications.}
\section{The art of sampling best circuits}\label{appndix:art_of_sampling}
The process of selecting optimal quantum circuits from those proposed by the reinforcement learning (RL) agent involves a filtering threshold defined as:
\begin{figure}[t!]
    \centering
    \includegraphics[width=0.9\columnwidth]{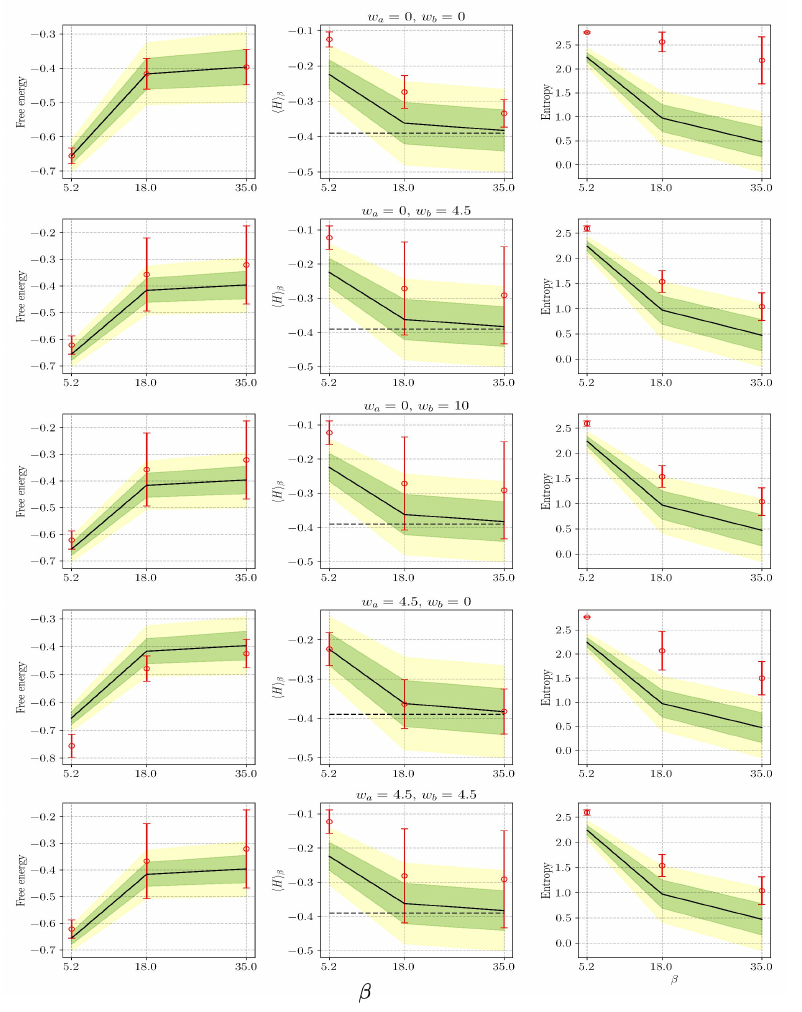}
    \caption{\small\textbf{Impact of the weights in the filtering threshold in Eq.~\ref{eq:filter_threshold} on the sampling of thermal states}. RL-agent proposes many possible structures of PQC$_2$ and it is important to find a tradeoff in the weight values to sample the best thermal state.}
    \label{fig:sampling_art1}
\end{figure}
\begin{equation}
\varepsilon = \Delta F + w_a \Delta \langle H\rangle_\beta + w_b \Delta S,
\end{equation}
where $\Delta F$, $\Delta \langle H\rangle_\beta$, and $\Delta S$ represent errors in free energy, Hamiltonian expectation, and entropy, respectively. Weights $w_a$ and $w_b$ adjust the relative importance of these errors.
The best-performing circuit is selected by: (1) Evaluating each proposed circuit using $\varepsilon$. (2) Selecting the circuit that minimizes $\varepsilon$.

\begin{figure}[h!]
    \centering
    \includegraphics[width=0.8\columnwidth]{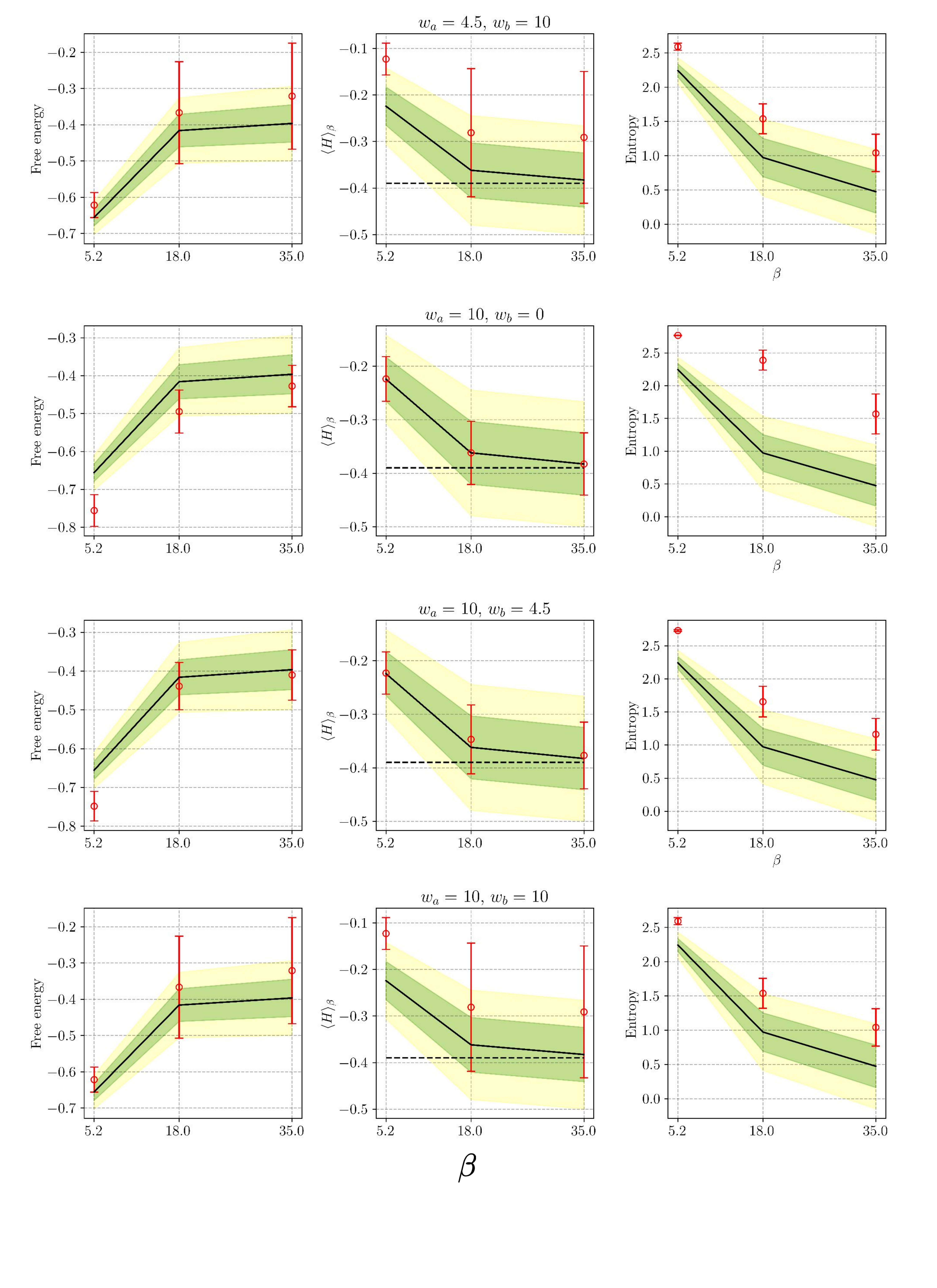}
    \caption{\small\textbf{Impact of a wider range of weights in the filtering threshold in Eq.~\ref{eq:filter_threshold} on the sampling of thermal states}. Continuation of the results in Fig.~\ref{fig:sampling_art2} for more weight values.}
    \label{fig:sampling_art2}
\end{figure}

Figures \ref{fig:sampling_art1} and \ref{fig:sampling_art2} illustrate the significant impact of weight selection on circuit sampling. These results emphasize the critical nature of careful postprocessing to achieve an optimal balance between free energy, Hamiltonian expectation value, and entropy of the thermal state.
The findings underscore the importance of thoughtful parameter selection in the filtering process, as it directly influences the quality of the sampled circuits. This approach allows for fine-tuning the selection criteria to meet specific performance requirements in quantum circuit design and optimization.

\end{document}